\documentclass[nopreprintline,3pt,final]{elsarticle}
\usepackage{stackengine}
\setstackEOL{\\}
\usepackage[pages=some,placement=bottom,color=black,opacity=1,scale=1,vshift=25pt]{background}

\usepackage[utf8]{inputenc}
\usepackage{microtype}
\usepackage[bottom]{footmisc}

\usepackage{framed,multirow}
\usepackage{subfigure}

\usepackage{lineno}

\usepackage{amssymb}  
\usepackage{amsthm}
\usepackage{amsmath}
\usepackage{latexsym}

\usepackage{booktabs}
\usepackage{longtable}
\usepackage{tabularx}
\usepackage{multirow}

\newcolumntype{C}{>{\centering\arraybackslash}X}
\newcolumntype{L}{>{\raggedright\arraybackslash}X}
\newcolumntype{R}{>{\raggedleft\arraybackslash}X}
\newcolumntype{P}[1]{>{\raggedright\arraybackslash}p{#1}}

\usepackage{url}
\usepackage{xcolor}
\usepackage{lineno}
\usepackage{hyperref}

\definecolor{newcolor}{rgb}{.8,.349,.1}

\definecolor{AN}{RGB}{0,0, 0}
\definecolor{HK}{RGB}{0, 0, 0}

\journal{Signal Processing: Image Communication}

\begin{document}
\BgThispage
\backgroundsetup{contents={\footnotesize%
    \Longstack{\textit{© 2020. Accepted for publication in Signal Processing: Image Communication.}\\
      \textit{This manuscript version is made available under the CC-BY-NC-ND 4.0 license:}\\
      \url{http://creativecommons.org/licenses/by-nc-nd/4.0/}}}}

\begin{frontmatter}

\title{\bf{BINet: a binary inpainting network for deep patch-based image compression}}

\author[1]{Andr\'e Nortje  \corref{cor1}}
\cortext[cor1]{Corresponding author: 
	\textit{e-mail:} \ttfamily{\bfseries{adnortje@gmail.com}} (Andr\'e Nortje)
}
\author[2]{Willie Brink}
\ead{wbrink@sun.ac.za}
\author[1]{Herman A. Engelbrecht}
\ead{hebrecht@sun.ac.za}
\author[1]{Herman Kamper}
\ead{kamperh@sun.ac.za}

\address[1]{Department of Electrical and Electronic Engineering}
\address[2]{Division of Applied Mathematics\\ 
Stellenbosch University, South Africa}

%% ABSTRACT
% 
% BINet Abstract
% 

\begin{abstract}
Recent deep learning models outperform standard lossy image compression codecs. However, applying these models on a patch-by-patch basis requires that each image patch be encoded and decoded independently. The influence from adjacent patches is therefore lost, leading to block artefacts at low bitrates. We propose the Binary Inpainting Network (BINet), an autoencoder framework which incorporates binary inpainting to reinstate interdependencies between adjacent patches, for improved patch-based compression of still images. When decoding a patch, BINet additionally uses the binarised encodings from surrounding patches to guide its reconstruction. In contrast to sequential inpainting methods where patches are decoded based on previous reconstructions, BINet operates directly on the binary codes of surrounding patches without access to the original or reconstructed image data. Encoding and decoding can therefore be performed in parallel. We demonstrate that BINet improves the compression quality of a competitive deep image codec across a range of compression levels.
\end{abstract}


\end{frontmatter}

%% INTRODUCTION
%
% BINet Introduction
%

% START INTRODUCTION
\section{Introduction}
\label{sec:intro}
% STANDARD APPROACH
Over 60\% of Internet byte content consists of still images~\cite{GoogleDevelopers2016}.
Efficient image compression is therefore essential in lowering transmission bandwidth and data storage costs. 
Lossy image compression is currently dominated by patch-based standard 
codecs such as JPEG~\cite{Wallace1991} and WebP~\cite{GoogleDevelopers2016}.
Patch-based encoding schemes are preferred to their full-resolution counterparts, 
as they are more memory efficient and are required by standard video codecs such as 
H.264/5 that rely on block motion estimation techniques~\cite{Richardson2010}.

% START RELATED WORK
Careful engineering has enabled standard image codecs 
to perform well in most settings.
But these codecs suffer from arduous hand-tuned parameterisation, 
which can be particularly sensitive 
to settings outside of the domain 
for which they were designed.
In contrast, deep neural networks are trained through
loss-driven end-to-end optimisation, 
and deep image compression models have been shown 
to outperform standard image codecs~\cite{Toderici2015,Balle2016,Toderici2017,Theis2017,Rippel2017,Agustsson2018,Johnston2018,Santurkar2018,Balle2018,Mentzer2018,Li2018, Lee2019}.
These deep approaches, although effective, are not optimised for patch-based encoding
since they use the full image content to steer compression.
Full image context is, unfortunately, 
not available for patch-based systems as each patch is encoded independently.
Patch-based encoding is 
therefore avoided in deep compression models~\cite{Toderici2015,Johnston2018}, 
as it may result in block artefacts at shallow bitrates.
To remedy this, we propose the Binary Inpainting Network (BINet) framework, 
which is inspired by research in image inpainting.

\begin{figure}[h]
        \centering
        \subfigure[Traditional]{
                % Full Context Region
                \includegraphics[width=.25\textwidth]{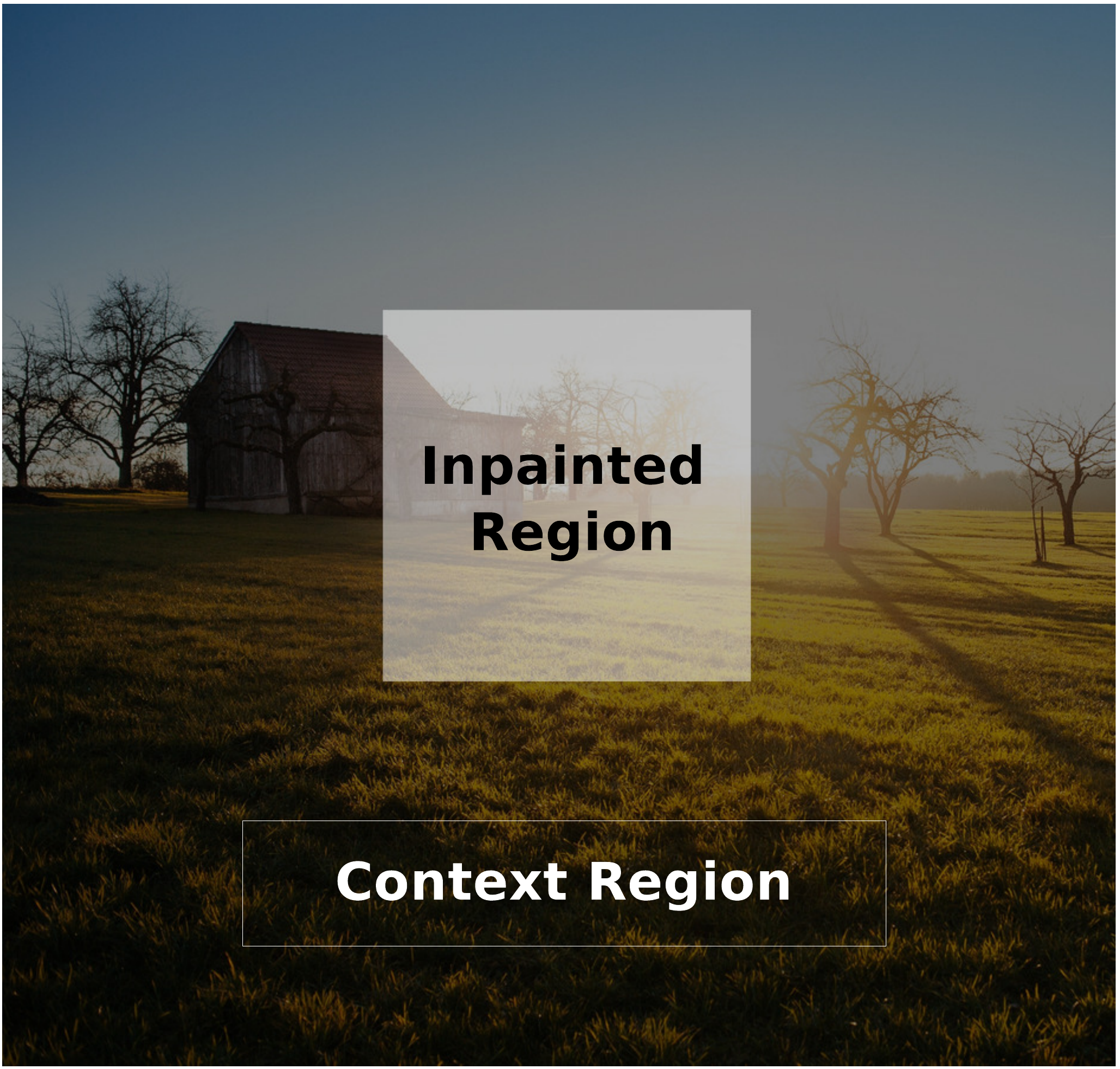}
		\label{fig:full_context}
        }
        \subfigure[BINet]{
                % BINet Context Region
                \includegraphics[width=.25\textwidth]{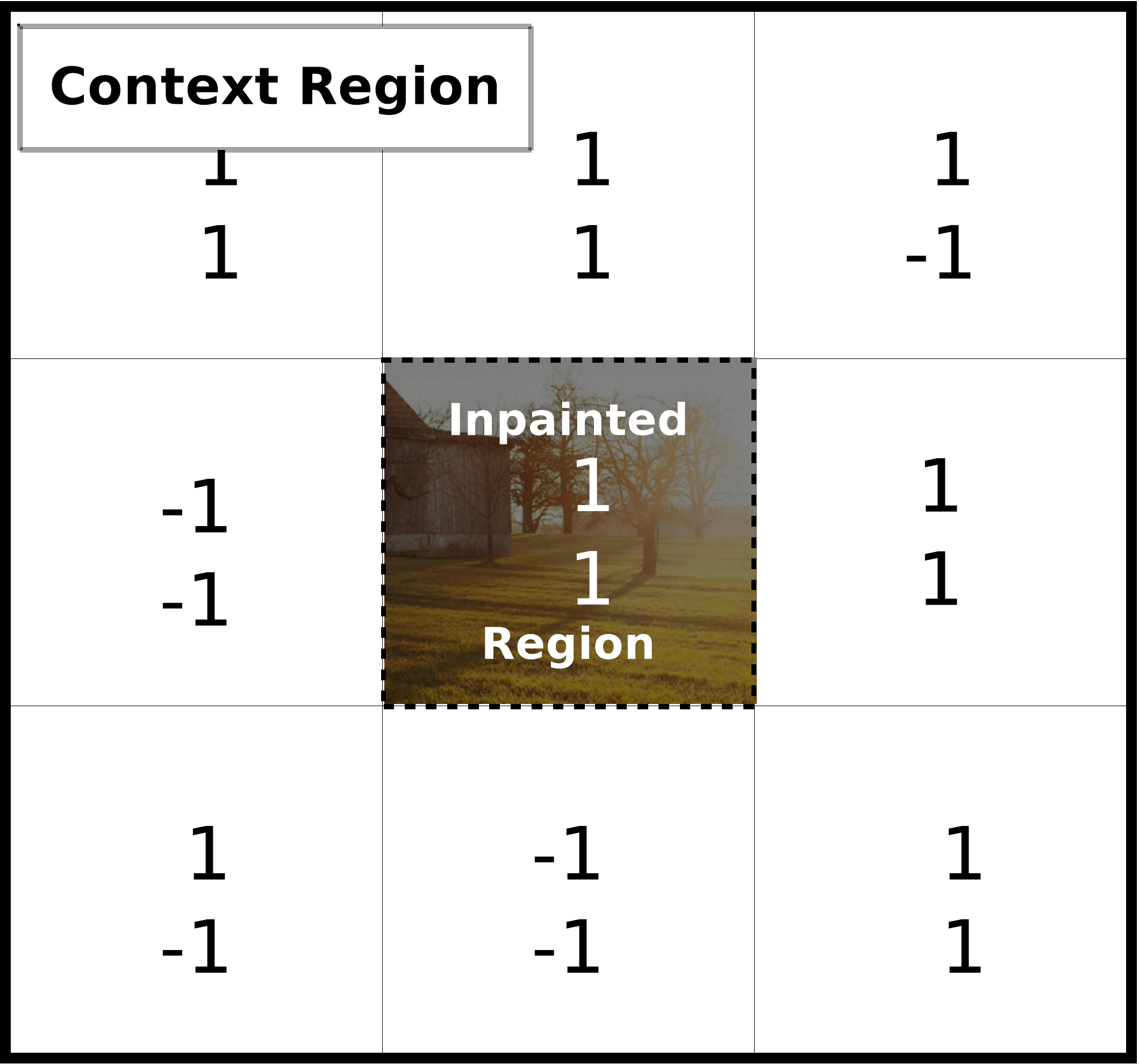}
                \label{fig:binet_context}
        }
        \subfigure[Sequential]{
                % Partial Context Region
                \includegraphics[width=.25\textwidth]{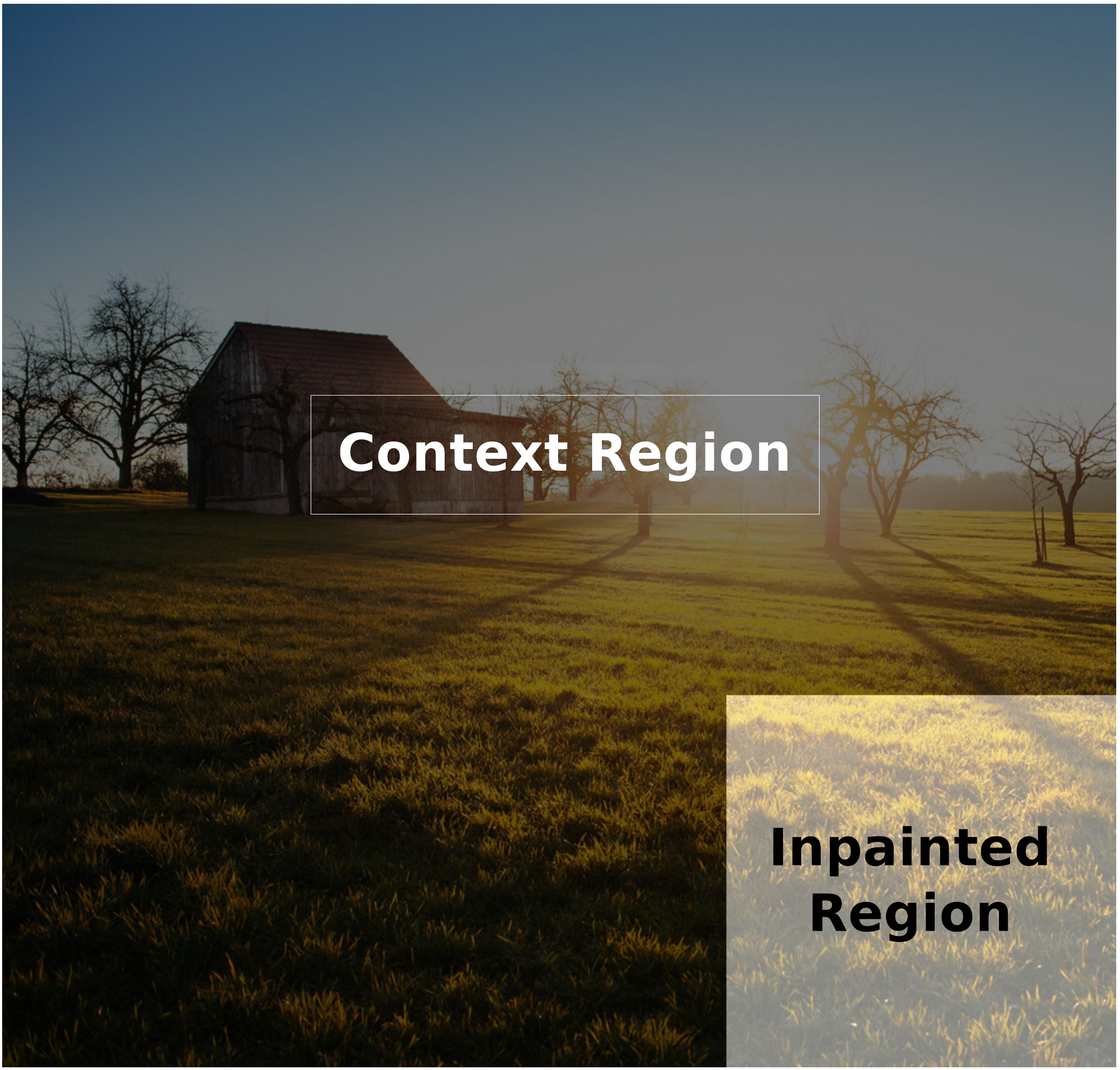}
		\label{fig:partial_context}
        }
        \caption{
                Context regions available to various inpainting models. In traditional inpainting (a), a masked-out region is predicted based solely on context from surrounding regions. In BINet (b), the middle region is decoded based on binary codes (indicated with the $-1$'s and $1$'s) of the patch of interest but also the binary codes from adjacent image patches. In sequential compression methods (c), patches are decoded in a pre-specified order; these reconstructed patches can then be used as the context region when decoding a next patch in the sequence.
        }
        \label{fig:context_regions}     
\end{figure}

\begin{figure}[b]

        \centering
        \includegraphics[width=.99\textwidth]{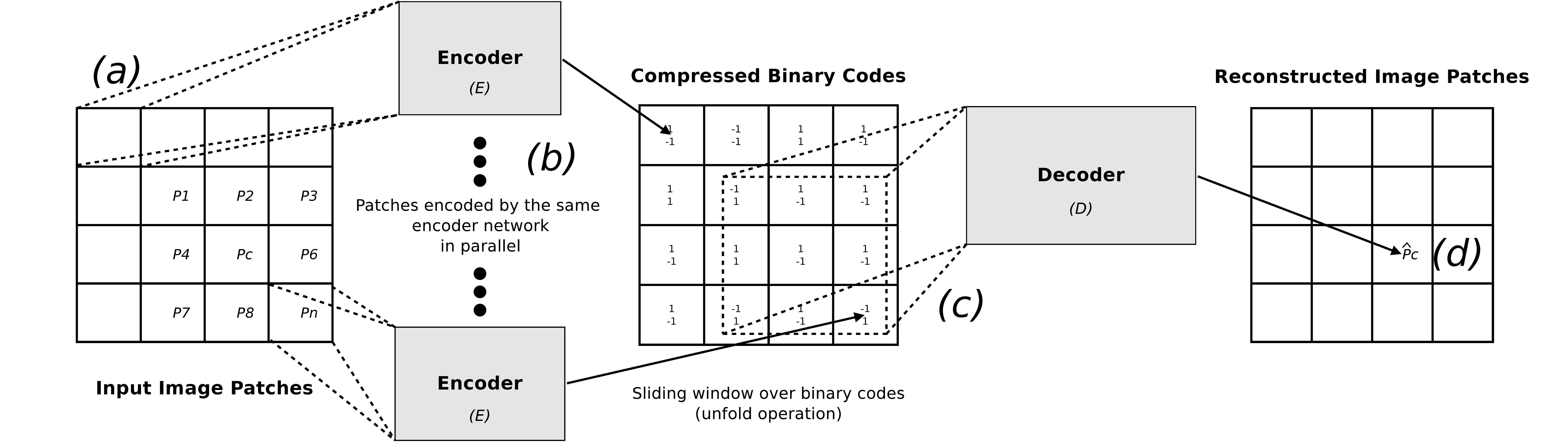}       
        \caption{
                The Binary Inpainting Network (BINet) framework.
                Compressed binary codes are illustrated here as 
                two bits ($-1$ or $1$) per patch.
        }
   
        \label{fig:bin_overview}
        
\end{figure}

% BINet VS OTHER INPAINTING STRATEGIES
Image inpainting involves reconstructing a masked-out image region
by using the surrounding pixels as context.
It is often used as an error-correction strategy to restore patches lost during transmission.
Traditional inpainting models, such as PixelCNN~\cite{Oord2016}, 
assume access to original pixel content; 
in Figure~\ref{fig:full_context}, 
the model would be asked to predict 
the shaded region in the middle,
given the surrounding context as input.
We extend this idea in order to perform patch-based image compression.
When decoding a particular patch, 
BINet incorporates the compressed binary codes from adjacent image patches as well as 
the current patch to reinstate relationships between separately encoded regions.
As depicted in Figure~\ref{fig:binet_context},
BINet therefore exploits encoded binary information from a full-context region as well as 
the patch being inpainted in order to formulate its prediction of the inpainted region. 
The overall approach is illustrated in Figure~\ref{fig:bin_overview}: 
BINet encodes patches as discrete binary codes using a single encoder. 
The decoder then reconstructs a particular centre patch by incorporating 
the binary codes of surrounding patches.
It therefore allows for parallel encoding and decoding 
of image patches aided by learned inpainting from a full binary context region.

In sequential compression techniques such as WebP~\cite{GoogleDevelopers2016}, 
linear combinations of previously reconstructed outputs are used 
when decoding a particular patch. 
\textcolor{AN}{This is similar to sequential patch-based inpainting~\cite{Baig2017,Fabian2019}}, 
as illustrated in Figure~\ref{fig:partial_context}, 
where previously decoded output from the model is treated as the context region 
and used to perform inpainting on the next patch.
In contrast to these approaches, 
BINet decodes a particular patch, not based on previous patch reconstructions, 
but based directly on the binary encodings of the surrounding patches.
Since it does not need to wait for surrounding patches to be decoded, 
BINet can decode all patches in parallel while still taking 
the full surrounding context into account.

BINet's encoder and decoder are trained 
jointly through end-to-end optimisation.
In contrast to~\cite{Baig2017}, where separate compression and inpainting networks are trained,
BINet builds inpainting directly into its decoder architecture 
and does not require training an additional inpainting network.
Our aim is to show that this approach allows spatial dependencies between patches to be re-instated from independently encoded patches, 
thereby advancing patch-based encoding in a neural compression model.

% PAPER OUTLINE
We proceed with a description of the BINet framework and
the formulation of a loss function for learning binary encodings that exploit 
spatial redundancy between neighbouring image patches.
BINet can be used with different types of encoder and decoder architectures, 
and in this work we specifically employ two competitive iterative decoding methods~\cite{Toderici2015, Toderici2017}, 
namely additive reconstruction (AR) and one-shot reconstruction (OSR).
We describe these specific instantiations of BINet in Section~\ref{sec:binet_arch}.
To show the benefit of incorporating inpainting, 
the BINet models are compared to convolutional AR 
and OSR models without inpainting.
Compression efficiency is evaluated quantitatively using the SSIM 
and PSNR image quality metrics.
We show that BINet performs better than the conventional AR and OSR approaches over the complete range of compression levels considered (Section~\ref{sec:binet_experiments}).
On the standard Kodak dataset~\cite{Kodak1999}, 
we show that the OSR variant of BINet consistently outperforms JPEG.
Although it falls short of outperforming WebP, 
we show qualitatively  that BINet produces 
smoother image reconstructions and 
is capable of more complex inpainting
than the sequential decoding methods used by WebP.
We released a full implementation of BINet online\footnote{\url{https://github.com/adnortje/binet}}.

% BINet Architecture Overview Diagram

% END INTRODUCTION

%% ARCHITECTURE
% START ARCHITECTURE
\section{Binary Inpainting Network (BINet)}
\label{sec:binet_arch}

% BINET ARCHITECTURAL OVERVIEW
\subsection{Architectural Overview}
\label{sec:binet_arch_overview}

BINet is a variation of a basic autoencoder~\cite{Hinton2006}.
Figure \ref{fig:bin_overview} shows BINet's encoding and decoding process.
It accepts as input a set of image patches, indicated by \textit{(a)} 
in the figure,  
that are reduced to low dimensional representations and binarised, 
as shown at \textit{(b)}.
Binarisation is required for digitally storing and/or transmitting 
a compressed version of an image~\cite{Richardson2010}.
As in~\cite{Toderici2015,Raiko2014} a stochastic binarisation function is 
used during training by adding uniform quantisation noise.
This allows us to backpropagate gradients through the 
binarisation layer in the encoder by copying the gradients from 
the first decoder operation to the penultimate encoder layer.
The decoder network at \textit{(c)}  is applied as a 
sliding window across the generated binary codes 
such that each image patch at \textit{(d)} 
is decoded using both its own binary code and the codes of 
adjacent patches that fall within a specific grid region. 
Intuitively, because the encoder and decoder networks are trained jointly, 
the decoder learns to inpaint from binary codes within its context region 
whilst the encoder learns to produce more compact codes that promote 
the inpainting performed by the decoder.
The same encoder network is applied to each individual image patch, meaning 
that encoding on multiple patches can be performed in parallel.
In principle any model can be used as the encoder and decoder in Figure \ref{fig:bin_overview}, 
which is why we refer to BINet as a framework.

As depicted in Figure \ref{fig:bin_overview}, the reconstruction
of a patch $P_c$ from its compressed representation can be formulated as
% High Level BINet Equation
\begin{equation}
 	\textcolor{AN}{\hat{P_c} = {D}({E}(P_1), {E}(P_2), \ldots, {E}({P_c}), \ldots, {E}(P_n)),}
 	\label{eq:bin_overview}	
\end{equation}
where $E(\cdot)$ and $D(\cdot)$ represent the encoder and decoder 
mappings shown at \textit{(b)} and \textit{(c)}, respectively. 
$P_1, P_2, \ldots, P_n$ represent the $n$ patches used as context
for predicting the centre patch $P_c$. 
The sliding window at the decoder can be implemented 
using unfold operations to maintain parallelisation, 
and takes the bits produced for $P_1, P_2, \ldots, P_n$ 
at \textit{(b)} 
as context to make the prediction $\hat{P}_c$.
Note that the same encoder network is applied to each of 
the input image patches individually and in parallel.
Edge regions of the binary codes are appropriately padded
so that the spatial resolution of the input image is maintained.
To learn how to inpaint, we use the $L_1$ loss:
\begin{equation}
 	L_\textrm{inpaint} = | P_c - \hat{P}_c | 
 	=  |P_{c} - \mathrm{Auto}(P_1, P_2, \ldots, P_c, \ldots, P_n)|,
 	\label{eq:inpaint_loss}
 \end{equation} 
where $\mathrm{Auto}(\cdot)$ is equivalent to $D(E(\cdot))$.

% BINET WITH AR & OSR 
\subsection{Progressive BINet Architectures}
\label{sec:binet_osr_ar_arch}

BINet can be used with different types of encoder and decoder networks, 
and here we consider two specific progressive architectures.

\subsubsection{Additive Reconstruction (AR)}
\begin{figure*}[!h]
        \begin{center}
                \includegraphics[width=.99\textwidth]{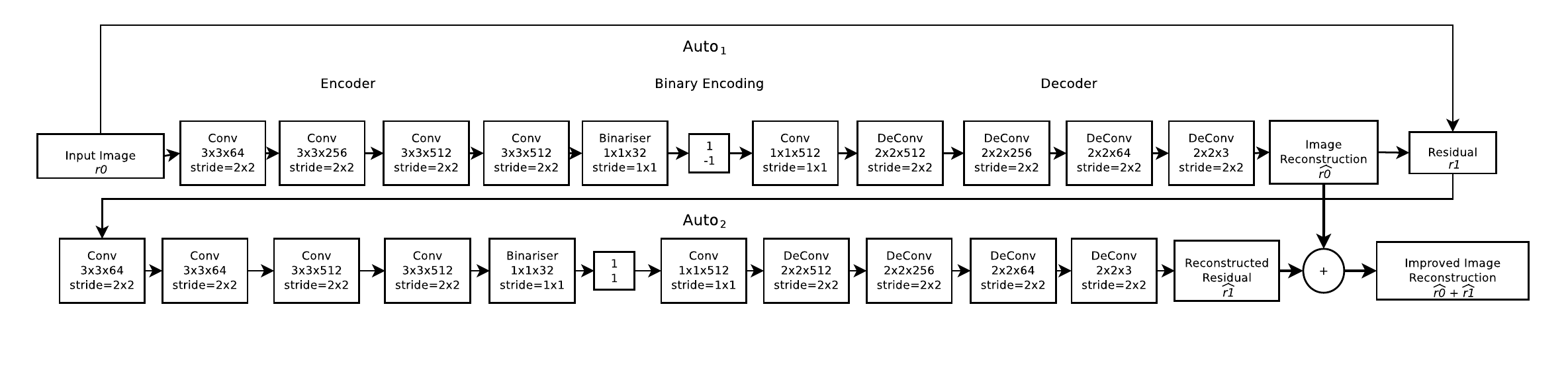}
                \vspace*{-10mm}
        \end{center}
        \caption{
                Two-iteration implementation of the Convolutional AR (ConvAR) baseline model.
        }
        \label{fig:conv_ar}
\end{figure*}

Additive reconstruction (AR) is widely used in traditional image codecs 
for variable bitrate encoding and progressive image enhancement~\cite{Wallace1991}.
Variable bitrate encoding entails assigning fewer 
bits to simpler image regions 
and vice versa, thereby reducing the overall bitrate on average.
Progressive image compression involves encoding an image such that it can be reconstructed 
at various quality levels as bits are received by the decoder. 
Using AR, this is achieved by transmitting the difference (the residual) 
between successive compression iterations 
and the original image such that the decoder can enhance its 
reconstruction by adding subsequently received residuals~\cite{Wallace1991}.
The AR process is shown in Figure~\ref{fig:conv_ar} 
and can be expressed mathematically as
%
% Additive Reconstruction Equation
\begin{equation}
         r_i = r_{i-1} - \mathrm{Auto}_i(r_{i-1}).
        \label{eq:ar_process}
\end{equation}
Each autoencoder stage,
$\mathrm{Auto}_i$, attempts to reconstruct the residual error 
$r_{i-1}$ from the previous stage,
with $r_0$ representing the original image~\cite{Toderici2015}. 
The reconstruction error $r_i$ is then passed to the following network iteration, 
which attempts to reconstruct it.
The final output image is obtained by summing over all the
residuals produced across multiple network stages.

% Baseline AR Figure

% BINet AR Figure
\begin{figure*}[!h]
        \begin{center}
                \includegraphics[width=.99\textwidth]{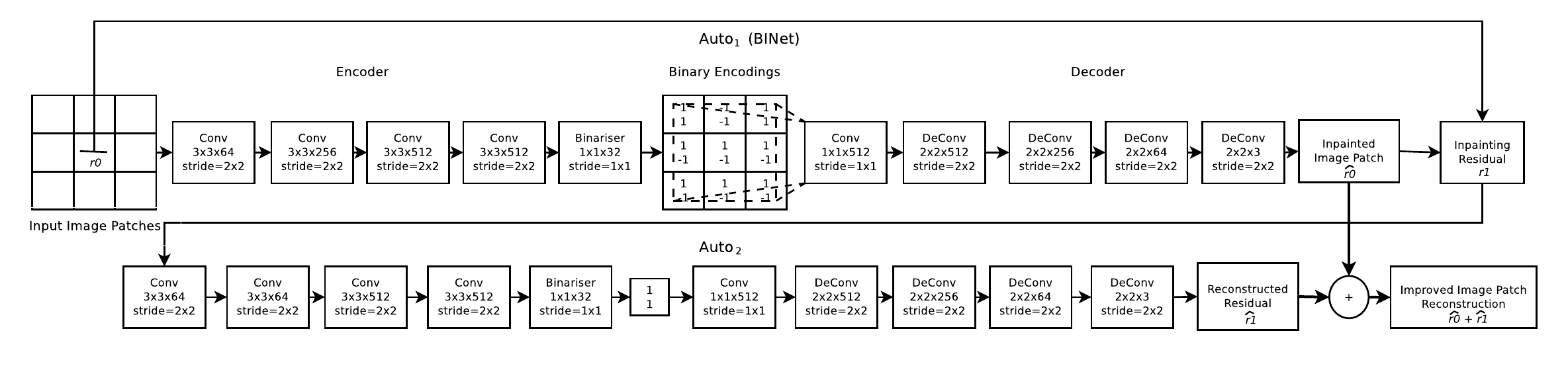}
                \vspace*{-5mm}
        \end{center}
        \caption{
                Two-iteration implementation of BINet with additive reconstruction (BINetAR).
                Binary inpainting is integrated into the first iteration of the ConvAR model in Figure~\ref{fig:conv_ar}.
        }
        \label{fig:bin_ar}      
\end{figure*}

\subsubsection{One-Shot Reconstruction (OSR)}
\begin{figure*}[!t]
        \begin{center}
                \includegraphics[width=.99\textwidth]{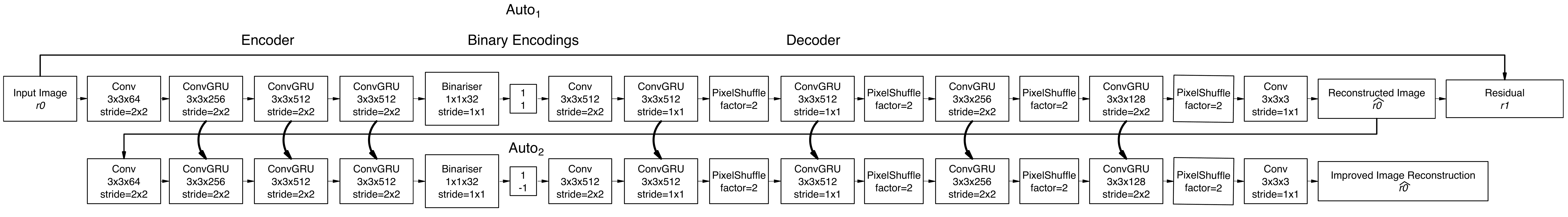}
        \end{center}
        \caption{
                \textcolor{AN}{Two-iteration implementation of the Convolutional GRU OSR (ConvGRU-OSR) baseline model.}
        }
        \label{fig:conv_osr}
\end{figure*}

One-shot reconstruction (OSR) 
is defined mathematically as follows~\cite{Toderici2017}: 
% One-Shot Reconstruction Equation
\begin{equation}
        r_{i} = r_{0} - \mathrm{Auto}_{i}(r_{i-1}).
        \label{eq:osr_process}
\end{equation}
% by equation~\eqref{eq:osr_process}.
Each iteration, $i$, accepts the previously incurred residual error, $r_{i-1}$, 
as input and uses it to reconstruct an improved quality approximation of the original image.
OSR differs from AR in that the original image is reconstructed 
at each network stage as opposed to the previous stage's residual.
This is achieved by recurrent links that 
propagate encoder and decoder state information.
The compression quality of the current iteration is thus influenced 
by relevant information from previous encodings and decodings 
that persist in the network's memory.
Figure~\ref{fig:conv_osr} illustrates the Convolutional Gated Recurrent Unit (GRU) OSR system~\cite{Toderici2017}, denoted as ConvGRU-OSR.

% Baseline OSR Figure

% BINet OSR Figure
\begin{figure*}[!t]
        \begin{center}
                \includegraphics[width=.99\textwidth]{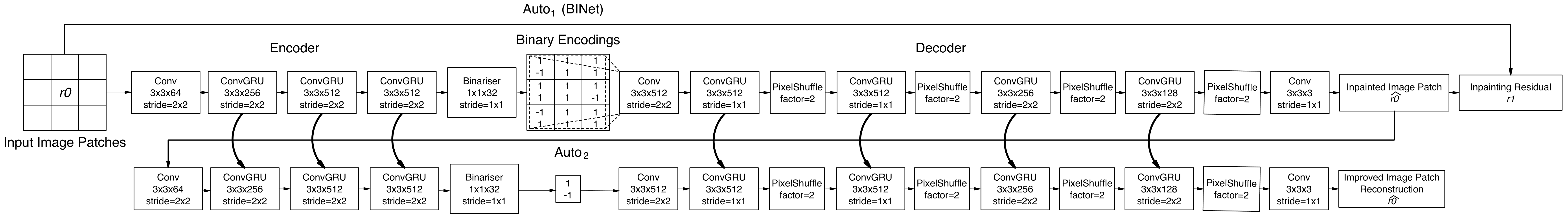}
        \end{center}
        \caption{
                \textcolor{AN}{Two-iteration implementation of BINet with one-shot reconstruction (BINetOSR).
                Binary inpainting is integrated into the first iteration of the ConvGRU-OSR model in Figure~\ref{fig:conv_osr}.}
        }
        \label{fig:bin_osr}     
\end{figure*}

%\subsubsection{BINet with Additive (AR) and One-Shot (OSR) Reconstruction}
\subsubsection{BINet with AR and OSR}
Both AR and OSR can be used naturally with BINet. 
As baselines, we use the progressive ConvAR~\cite{Toderici2015,Baig2017} 
and ConvGRU-OSR~\cite{Toderici2017} networks shown in Figures~\ref{fig:conv_ar} and~\ref{fig:conv_osr}, respectively.
The reconstruction of an image patch $P$ for a single iteration of
these models can be written as:
%
% Patch Reconstruction Equation (No Inpaint)
\begin{equation}
	\hat{P} = \mathrm{Auto_1}(P) = {D_1}({E_1}(P)).
\end{equation}
Their patch reconstruction are therefore
based on the encoding of a single input patch $P$.
In other words, they do not incorporate inpainting to aid compression.
The training loss for both the ConvAR and ConvGRU-OSR 
baselines can be expressed as
% Additive Reconstruction Loss
\begin{equation}
	L_\textrm{baseline} = \sum\limits_{i=1}^{I}|r_i|,
	\label{eq:baseline_loss}
\end{equation}
where $I$ refers to the number of reconstruction iterations 
(Figures~\ref{fig:conv_ar} and~\ref{fig:conv_osr} show only two iterations,
but typically more are used).

The BINet framework is incorporated into ConvAR and ConvGRU-OSR 
by including learned binary inpainting at the first iteration,
as shown in Figures~\ref{fig:bin_ar} and~\ref{fig:bin_osr}.
Later iterations encode the residual error incurred 
by this initial inpainting prediction. 
We only include inpainting at the first iteration,
as intuitively this stage encodes details that contain 
the most spatial redundancy compared to later stages whose 
purpose is to encode finer and less correlated patch details.\footnote{Future work may focus on ways of including binary inpainting at later network stages.}
Our goal is to show the benefit of this binary inpainting strategy.

The encoding process of BINetAR (Figure~\ref{fig:bin_ar}) 
and BINetOSR (Figure~\ref{fig:bin_osr}) can be expressed as in 
equations~\eqref{eq:ar_process} and~\eqref{eq:osr_process}, 
where $r_0$ again represents the original input image patch while $r_1$ 
is the initial iteration's 
inpainting loss given in equation~\eqref{eq:inpaint_loss}. 
An $I$-iteration implementation of BINet
with either AR or OSR is trained to optimise the loss: 
% Equation : BINet Loss
\begin{equation}
 	L_\textrm{BINet} = L_\textrm{inpaint} + \sum\limits_{i=2}^{I}|r_i|.
	\label{eq:binet_loss}
\end{equation}
%
% END ARCHITECTURE

%% TRAIN
% START EXPERIMENTAL SETUP AND PROCEDURE

\section{Experimental Setup}

\subsection{Data and Training Procedure}
\label{sec:train_procedure}
The models discussed in Section~\ref{sec:binet_arch} are trained on 
the CLIC Compression Challenge Professional Dataset~\cite{Freeman2018}, 
which is pre-partitioned into training, validation and test sets.
Each set contains a variety of professionally captured high resolution natural images, 
saved in lossless PNG format to prevent the learning of 
compression artefacts introduced by lossy codecs.

% Training Procedure 
The loss functions in equations~\eqref{eq:baseline_loss} 
and \eqref{eq:binet_loss} are used to train $I=16$ iteration 
implementations of the baseline (ConvAR, ConvGRU-OSR) 
and BINet (BINetAR, BINetOSR) systems, respectively. 
All models are trained to encode and 
reconstruct randomly cropped $32\times32$ image patches.
Following the approach in~\cite{Toderici2015}, 
the networks are constrained such that each autoencoder 
stage contributes  0.125 bits per pixel (bpp) 
to the overall compression of an input image patch.
During training, BINet encodes nine directly adjacent image patches independently 
and reconstructs the central patch region based on 
the binary codes produced for the nine patches.
Training patches are randomly cropped from the images in the training set 
at every epoch while centre cropping is used on images in the validation set 
to ensure that the validation losses for the BINet 
and baseline models are directly comparable across epochs.
Image patches used during training are batched into groups of 32 
and normalised such that pixel values fall in the range $[-1, 1]$.
Models are trained for 15\,000 epochs and 
early stopping is employed based on the validation loss.\footnote{
For the preliminary analyses in Section~\ref{sec:prelim_analysis} we stop training at 5\,000 epochs.}
We use Adam optimisation~\cite{Kingma2014} 
with an initial learning rate of 0.0001.  
The learning rate is decayed by a factor of 2 at 
epochs 3\,000, 10\,000 and 14\,000.

\subsection{Evaluation Procedure}
% Quality Metrics
Quantifying image quality in a way that aligns with the 
subjective nature of the human visual system is difficult. 
A subjective assessment using surveys on humans can be slow 
and prone to viewer bias, 
and may garner results that are not easily reproducible.
Objective algorithms have therefore become the norm in assessing 
image compression models~\cite{Richardson2010}. 
We use two standard objective image evaluation metrics:
Peak Signal to Noise Ratio (PSNR) 
and Structural SIMilarity index (SSIM)~\cite{Wang2004}. 
PSNR and SSIM measure the degree to which an image reconstruction 
corresponds to the original image. 
In both cases a higher score implies greater fidelity. 
SSIM falls within $[-1,1]$ while PSNR (usually expressed in dB) 
can be any real value. 
We follow the procedure recommended in~\cite{Wang2004} 
when calculating the SSIM of a compressed image: 
the final SSIM score is obtained by applying the SSIM index 
over smaller ($11\times11$) pixel regions in a convolutional 
manner on a per-channel basis, and averaging the results.
We use $K_1=0.01$, $K_2=0.03$, and $\sigma = 1.5$,
with a Gaussian weighting process, as in~\cite{Wang2004}. 

% Evaluation Procedure
For evaluation, each image is resized to $320\times224$ pixels 
such that evaluation image dimensions are cleanly divisible by 
the chosen $32\times32$ patch size.\footnote{We also ran tests on full unscaled images, 
and found that trends were exactly the same as 
when images are resized in this way,
due to the models always compressing a fixed patch size
irrespective of input image dimensions.}
Images are then partitioned into $32\times32$ pixel patches and encoded, 
and quality scores are calculated on and averaged across the reassembled images.
The performance of BINet is contrasted to that of the baseline systems at various bit depths 
in order to gauge the effectiveness of incorporating the proposed 
binary inpainting framework across different operating points.
Additionally, we perform various 
preliminary analyses on validation data 
to further illustrate BINet's capabilities.

% END EXPERIMENTAL SETUP AND PROCEDURE

%% EXPERIMENTS
% START BINET EXPERIMENTS
% 18247717

\section{Experiments}
\label{sec:binet_experiments}

We first perform a preliminary analysis on development data 
to better understand the properties of BINet 
and the benefit of binary inpainting as opposed to conventional 
sequential inpainting techniques. 
We then turn to quantitative analyses on test data 
where BINet is compared to the baseline neural compression models 
as well as standard image compression codecs.

\subsection{Preliminary Analysis}
\label{sec:prelim_analysis}

\subsubsection{Is Inpainting from Binary Codes Possible?}
In order to assess qualitatively whether inpainting of image patches 
from compressed binary codes is possible, 
a 1-iteration implementation of BINetAR (0.125 bpp) is trained to explicitly 
predict the pixel content of an unknown $32\times32$ patch region located at 
the centre of a $96\times96$ pixel grid. 
This version of BINet is purposefully altered such that it masks bits
pertaining to the central patch region, 
i.e.\ the context region available to the decoder matches that of Figure~\ref{fig:full_context}.
This forces the network to become fully reliant on the binary encodings 
of surrounding patches when predicting the central patch's pixel content.

\begin{figure}[!h]
        \centering
                \includegraphics[width=.45\textwidth]{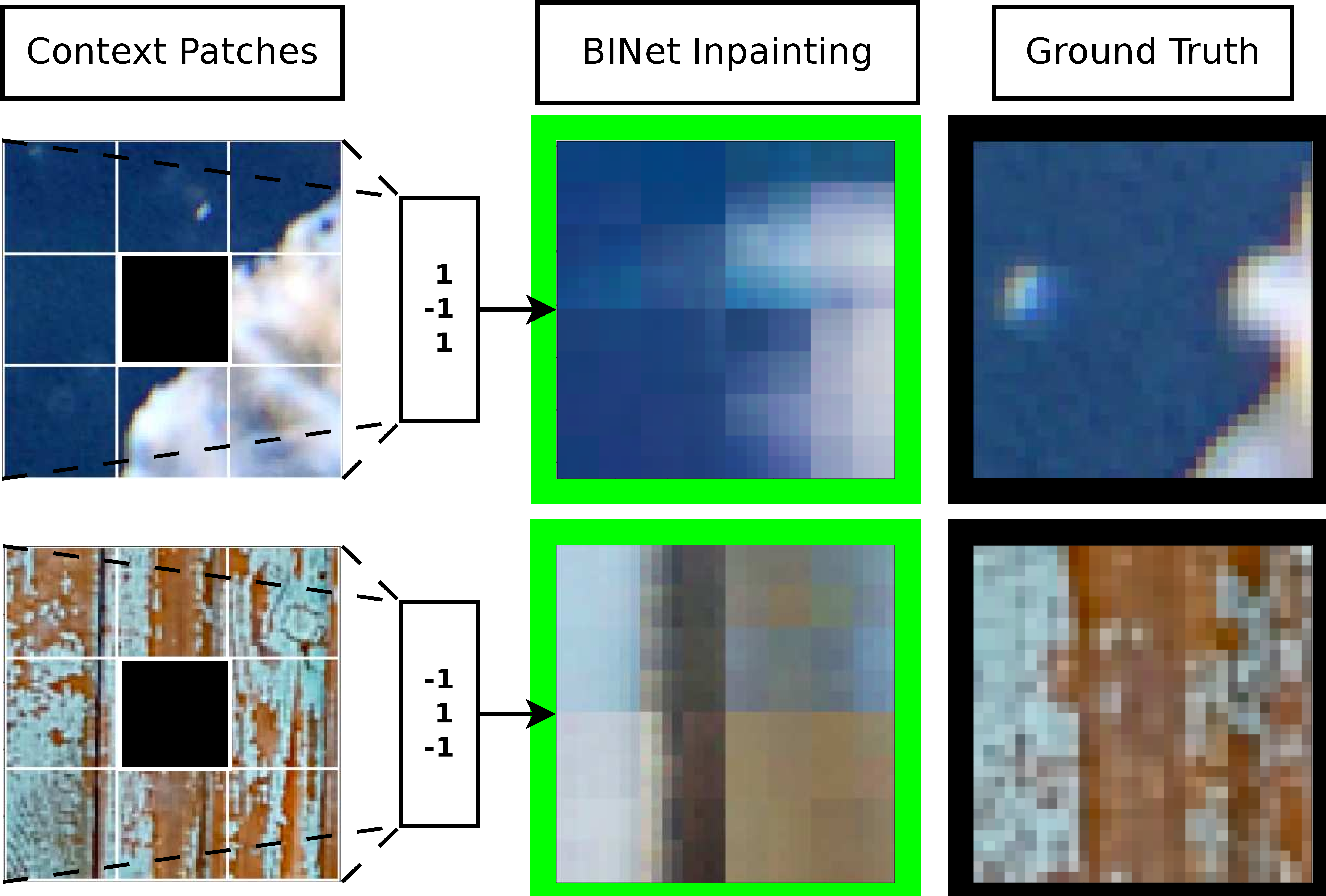}
        \caption{
                Inpaintings performed by masked BINet.
        }
        \label{fig:inpaint_vis}
\end{figure}

\begin{figure}[!h]
        \centering
                \includegraphics[width=.9\textwidth]{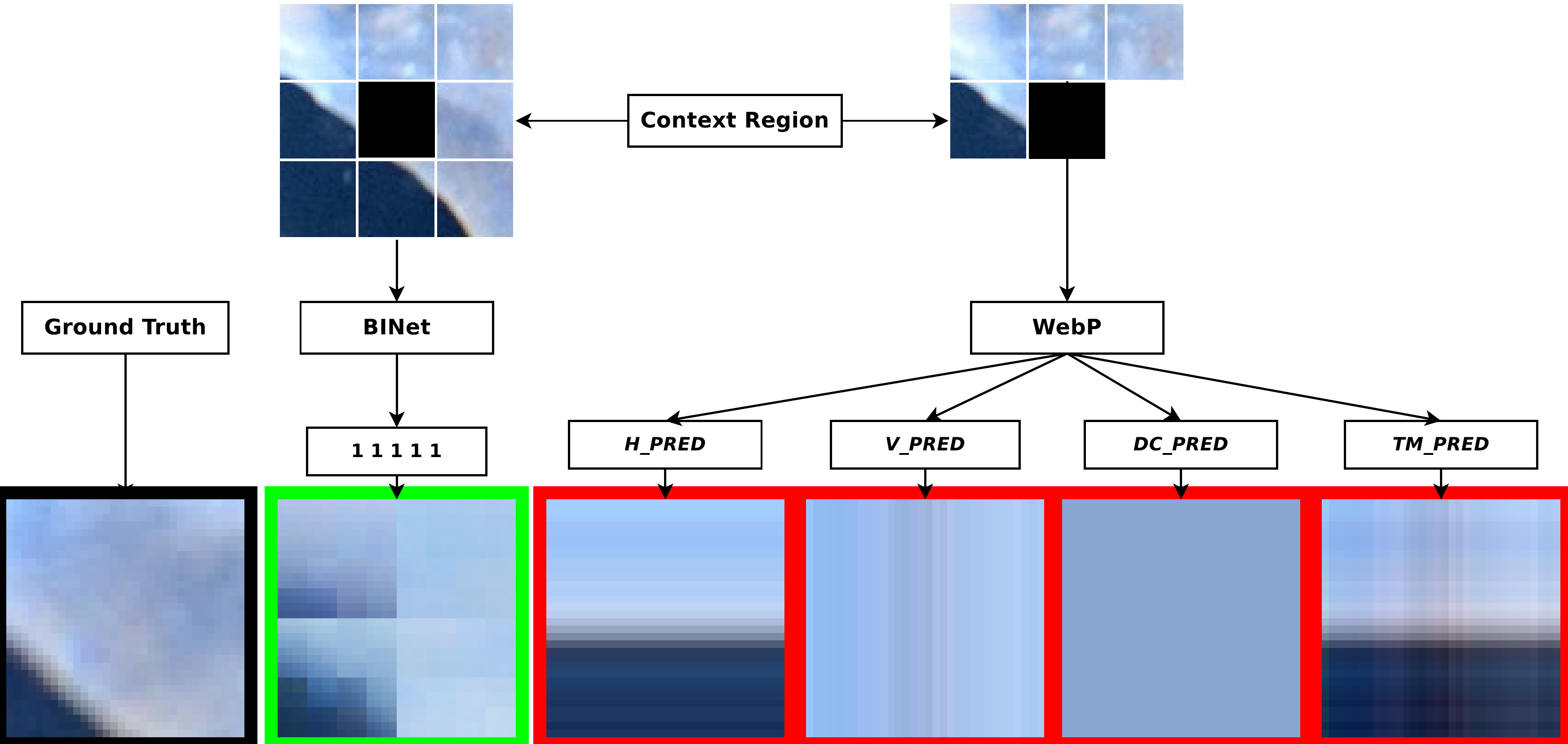}
        \caption{
                A comparison of the inpainting performed by masked BINet and WebP.
        }
        \label{fig:binet_vs_webp_vis}
        %\vspace{-3mm}
\end{figure}

Figure~\ref{fig:inpaint_vis} 
demonstrates the inpainting capabilities of this masked BINet, 
and indicates that it is able to predict a basis for an unknown patch using the compressed 
binary codes of its nearest neighbours.
Figure~\ref{fig:binet_vs_webp_vis} compares inpaintings 
from BINet (green border) and WebP (red border).
The four main modes used by WebP to sequentially predict a patch region 
are included in the diagram and abbreviated as in~\cite{GoogleDevelopers2016}.
The modes either average (DC\_PRED), directly copy (H\_PRED, V\_PRED), 
or linearly combine (TM\_PRED) pixels from previously decoded patches.
The figure shows that the inpaintings produced
by BINet resemble the ground truth patches (black border) more closely than those of WebP.

%
% Figure : Masked BINet Inpainting
%

% Figure : Masked BINet vs. WebP
%

%
% Figure : SINet
\begin{figure}[!h]

	\begin{center}
		\includegraphics[width=.99\textwidth]{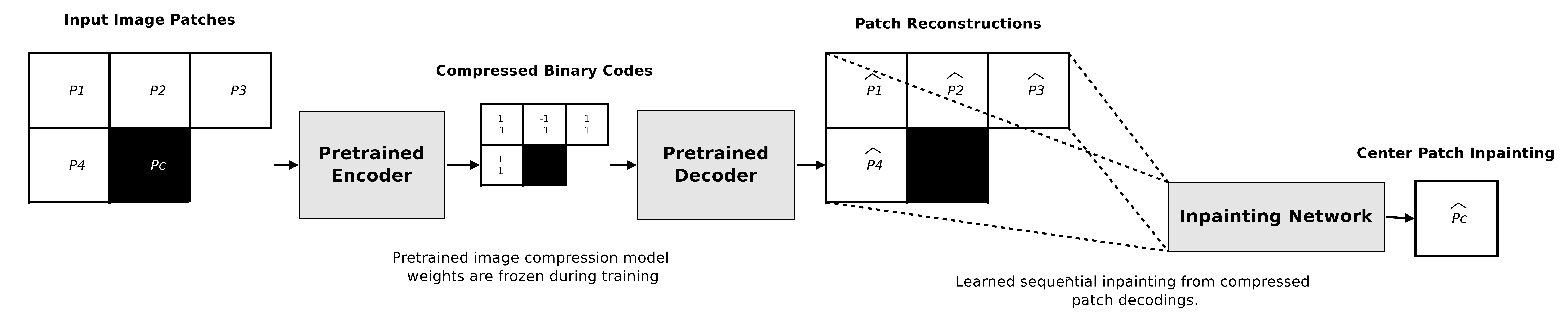}
		\vspace*{-5mm}		
	\end{center}
	
   	\caption{
   		The Sequential Inpainting Network (SINet).
	}
   
	\label{fig:seq_net}
	
\end{figure}

\subsubsection{Is Full-Context Binary Inpainting Superior to Sequential Inpainting?}
\begin{figure}[!h]
	\begin{center}	
		\includegraphics[width=.6\textwidth]{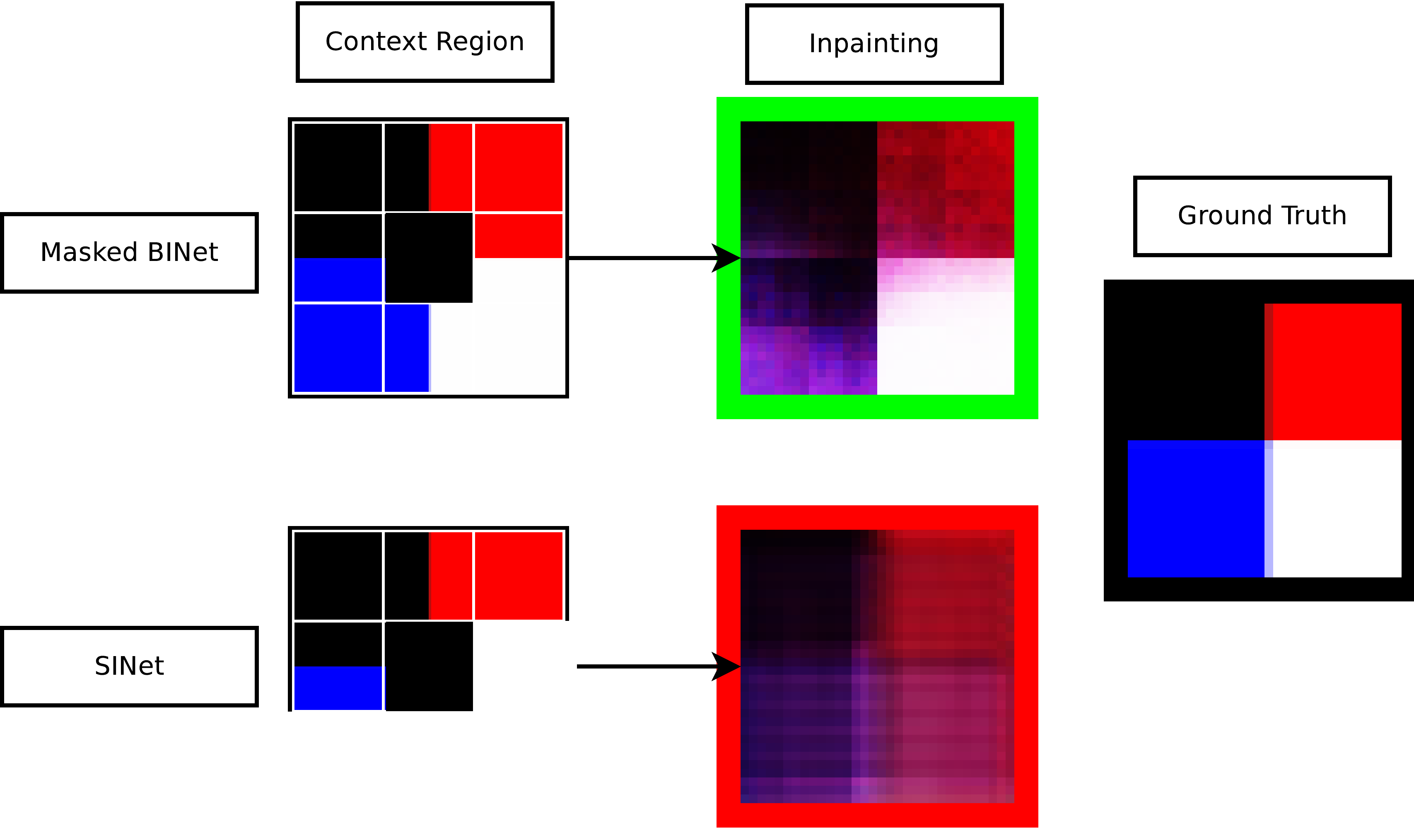}		
	\end{center}
   	\caption{
   		Comparison of inpainting performed by masked BINet 
   		and SINet given an artificial $32\times32$ image patch.
	}
	\label{fig:binet_vs_sinet}
\end{figure}

In this experiment we compare full-context binary inpainting to 
the sequential inpainting scheme proposed by~\cite{Baig2017}.
A masked 1-iteration realisation of BINetAR is pitted against the
Sequential Inpainting Network (SINet) in Figure~\ref{fig:seq_net}.
SINet consists of a pre-trained image compression model (ConvAR with $I=1$, bpp $=0.125$) 
coupled to an inpainting network (ConvAR decoder).
SINet's inpainting network is trained to sequentially predict the central patch $P_c$ 
from previously decoded patches such that its context region is like that 
of Figure~\ref{fig:partial_context}.
Table~\ref{tab:binet_vs_sinet} compares the
average SSIM 
and PSNR scores achieved by BINet and SINet on the validation set.
BINet's full-context binary inpainting mechanism leads 
to a 6\% improvement in SSIM 
and a 11\% increase in PSNR relative to the partial-context 
sequential inpainting performed by SINet.
Figure~\ref{fig:binet_vs_sinet} illustrates how BINet's ability to harness pixel 
content from a full context region aids its inpainting ability.
BINet (green border) correctly identifies that the lower right-hand corner of its inpainting should be white, 
whereas SINet (red border) is oblivious to this due to its limited context region.
Importantly, BINet has a major additional benefit in that it can be parallelised, since reconstruction of a particular patch is not performed based on previously decoded patches but rather directly on the binary codes of all 
surrounding patches.

%
% Figure : BINet vs SINet
%

% Table: BINet vs. SINet
\begin{table}[!h]
	\caption{Averaged masked BINet and SINet SSIM and PSNR scores for $32\times32$ image patch inpaintings}
	\begin{center}
	% \small
		\begin{tabularx}{.4\linewidth}{lCC}
			\toprule
			Model & SSIM & PSNR \\
			\midrule
				SINet  & $0.47$ & $19.85$ \\
				BINet  & $\mathbf{0.50}$ &$\mathbf{22.08}$  \\
			\bottomrule
		\end{tabularx}
	\end{center}
\label{tab:binet_vs_sinet}
\end{table}

\subsubsection{Does Inpainting Improve Compression using a Single Iteration?}
\begin{figure}[h]
\vspace{-5mm}
        \centering
	\subfigure[Training Loss]{
		\includegraphics[width=.4\textwidth]{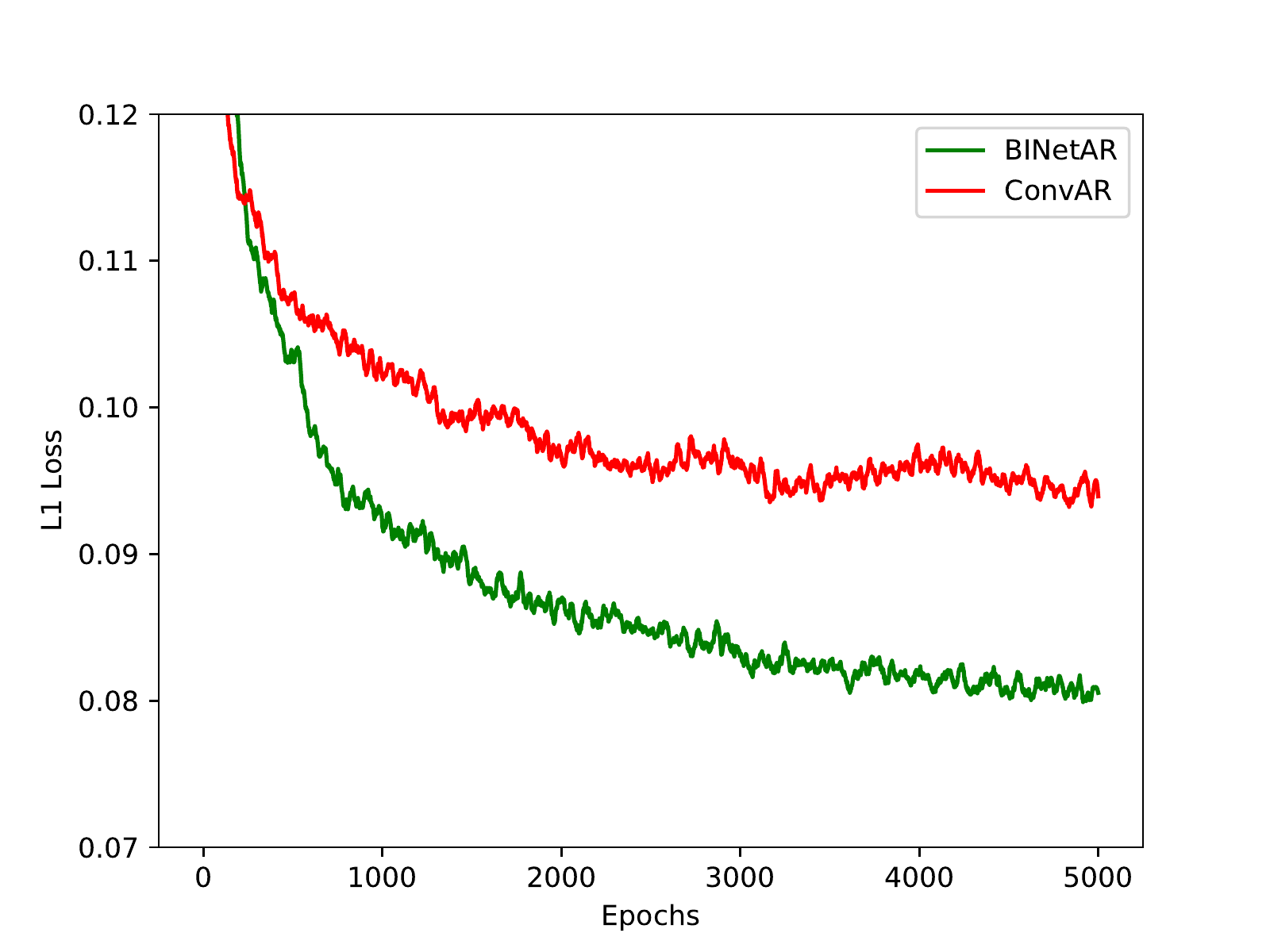}
	}
	\subfigure[Validation Loss]{
		\includegraphics[width=.4\textwidth]{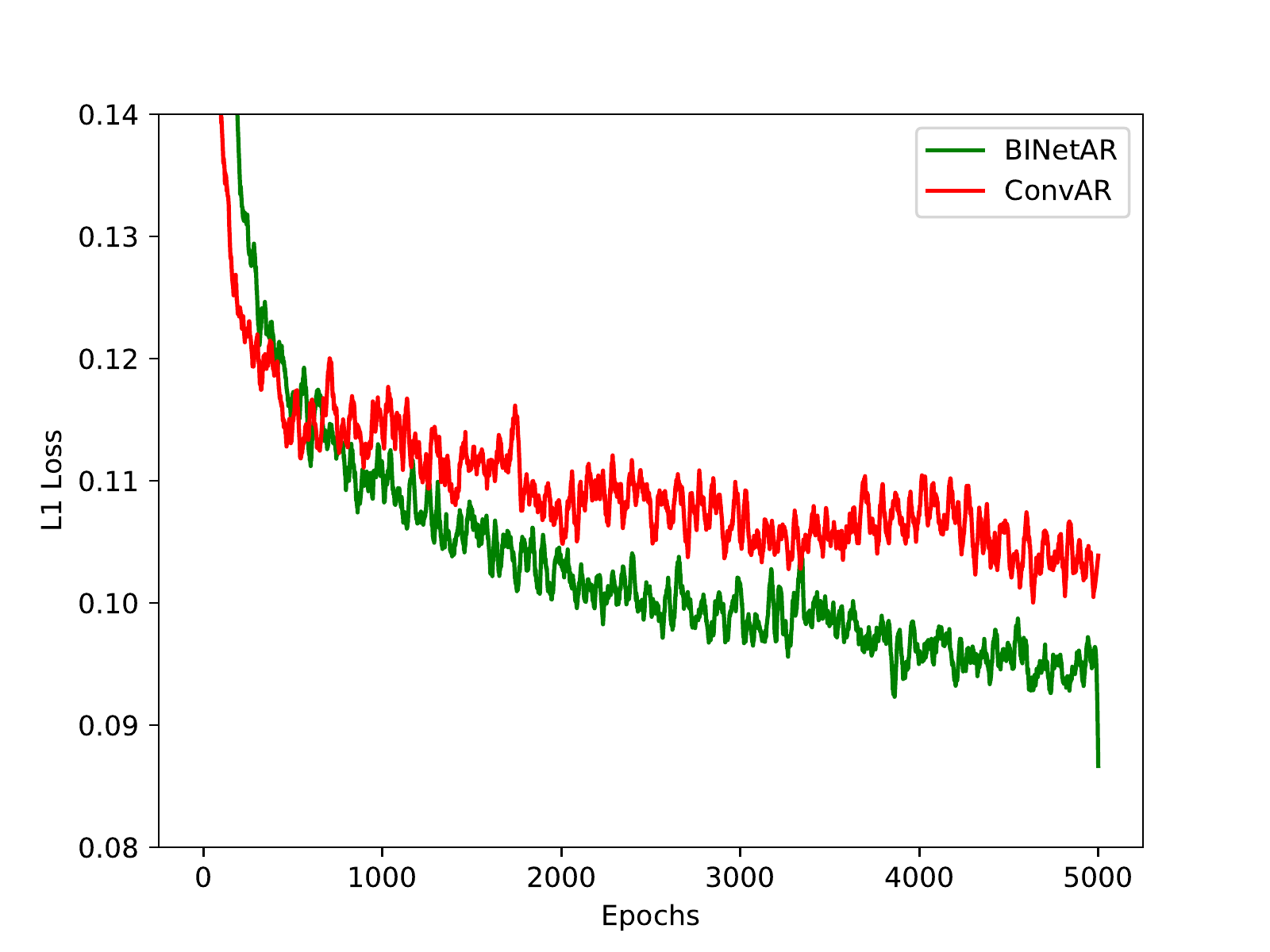}
	}
        \caption{
                Training and validation losses for 1-iteration implementations of BINetAR and the ConvAR baseline.
        }
        \label{fig:train_valid_loss1}
\end{figure}

\begin{figure}[!h]
	\centering
		\includegraphics[width=.85\textwidth]{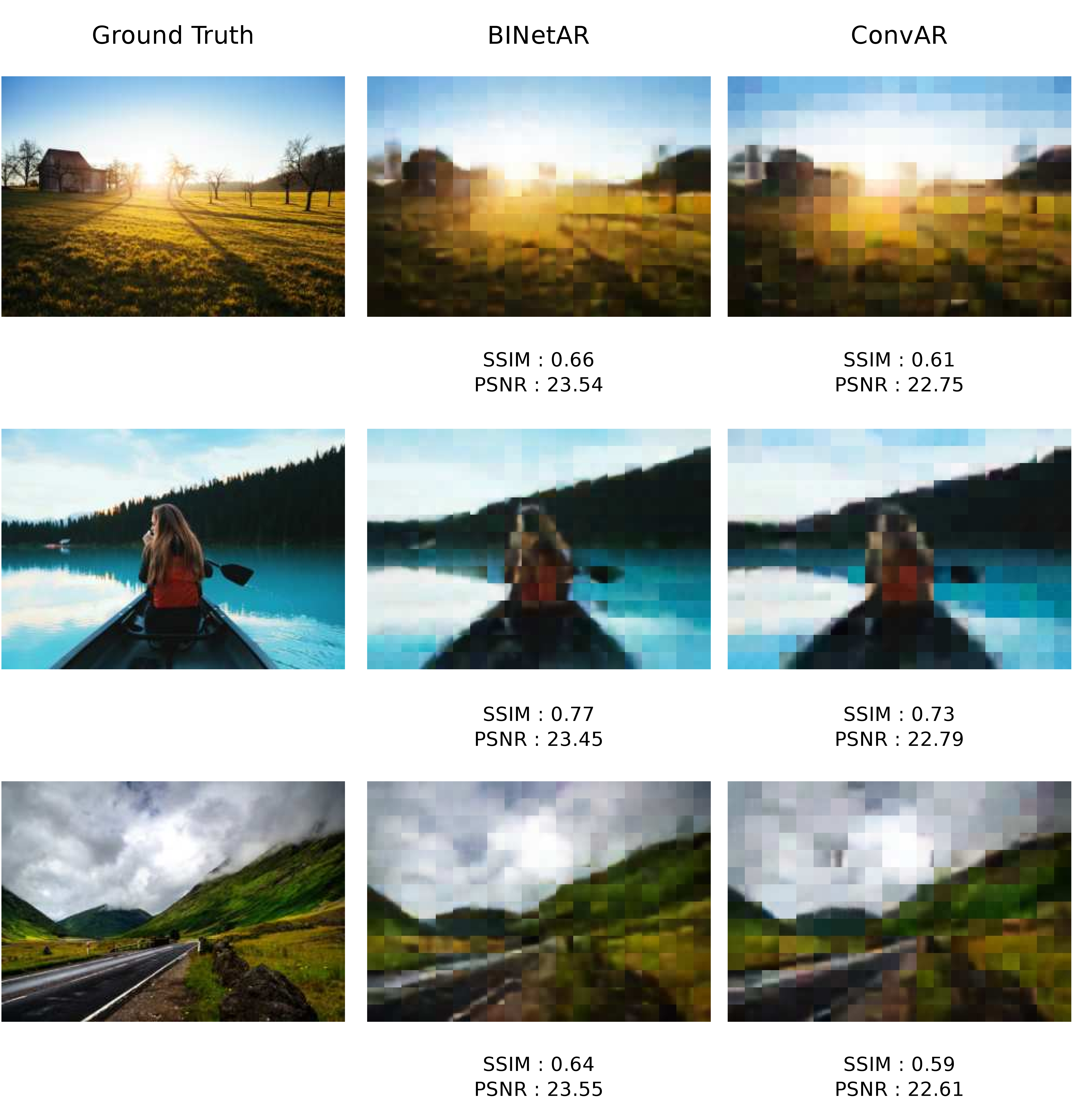}
	\caption{
		Full image reconstructions for 1-iteration implementations of BINetAR and the ConvAR baseline.
	}
	\label{fig:binet1_pics}	
\end{figure}

To determine if teaching a model to inpaint from binary codes aids its compression capabilities, 
1-iteration (0.125 bpp) implementations of BINetAR and the baseline ConvAR are pitted against each other.  
Figure~\ref{fig:train_valid_loss1} demonstrates 
how BINetAR outperforms ConvAR quantitatively in terms of training and validation loss. 
Losses represent the mean error between the ground truth and predicted patches 
and are indicative of the quality of the model's patch reconstructions.
Figure~\ref{fig:binet1_pics}
shows an assortment of images encoded by BINetAR and ConvAR. 
Note that in each case BINetAR produces images with a higher perceptual fidelity 
than ConvAR, according to the SSIM and PSNR scores achieved by its reconstructions. 
The images produced by BINetAR are qualitatively smoother 
than those of ConvAR at equally low bitrates, 
making BINetAR better suited for patch-based compression. 
The improved smoothness can be attributed to BINetAR's decoder which learns to 
constrain a patch to match its surroundings.
All the images used here are from the CLIC validation set~\cite{Freeman2018}.

%
% Figure : BINetAR vs ConvAR 1-Iteration Train and Validation Losses

\textcolor{AN}{\subsubsection{Does Inpainting Improve a Model's Ability to Adapt to Different Input Patch Sizes?}
Here we determine the influence of using different encoding patch-sizes on the compression quality of 1-iteration implementations of BINetAR and the ConvAR baseline model. 
Table~\ref{tab:patch_size_table} shows that BINetAR outperforms the baseline model for all the input patch-sizes considered. 
Interestingly, for smaller patch sizes the inclusion of binary inpainting improves compression quality, 
which seems sensible as smaller patches are less complex and easier to inpaint. 
Note that both models are trained to compress $32\times32$ patches (see Section~\ref{sec:train_procedure}), and are not retrained for this experiment. Table~\ref{tab:patch_size_table} therefore indicates that binary inpainting can improve a model's ability to adapt to different input sizes.
}
% Table : BINetAR vs. ConvAR with Different Patch Sizes
\begin{table}[!h]
\caption{
	BINetAR vs.\ ConvAR: SSIM and PSNR scores achieved when using different encoding patch sizes for the compression of $320\times320$ images to 0.125 bits-per-pixel (bpp).
}
% \vspace{-3mm}
	\begin{center}
	% \small
		\begin{tabularx}{.95\linewidth}{@{}lCCCCCCCC@{}}
			\toprule
			\multirow{2}{*}{Model} &
			\multicolumn{3}{c}{SSIM} & \multicolumn{3}{c}{PSNR} \\
			\cmidrule{2-7}
				& \small{$16\times16$} & \small{$32\times32$} & \small{$64\times64$} 
				& \small{$16\times16$} & \small{$32\times32$} & \small{$64\times64$} \\
			\midrule
				ConvAR & $0.587$ & $0.587$  & $0.587$ & $22.475$ & $22.474$ & $22.470$ \\
				BINetAR & $\mathbf{0.644}$ & $\mathbf{0.636}$ & $\mathbf{0.628}$ & $\mathbf{23.514}$ & $\mathbf{23.278}$ & $\mathbf{23.022}$ \\
			\bottomrule
		\end{tabularx}
	\end{center}
\vspace{-3mm}
\label{tab:patch_size_table}
\end{table}

%
% Figure : BINetAR vs ConvAR 1-Iteration
%

% END BINet EXPERIMENTS

%% EVALUATION
% START BINet EVALUATION
% 18247717

\subsection{Quantitative Analysis: BINetAR vs.\ ConvAR}
\label{sec:binet_ar_eval}

We now turn to quantitative analyses on the CLIC test set~\cite{Freeman2018}.
We train 16-iteration implementations of BINetAR and ConvAR to assess the 
effect of incorporating a single inpainting stage on the performance of an AR model.
We first consider reconstruction of single patches (an intrinsic measure) 
and then consider the more realistic evaluation on full images.

\subsubsection{Patch Reconstruction}
\begin{figure}[!h]
	\centering
	%\vspace{-5mm}
	\subfigure[SSIM]{
		\includegraphics[width=.45\textwidth]{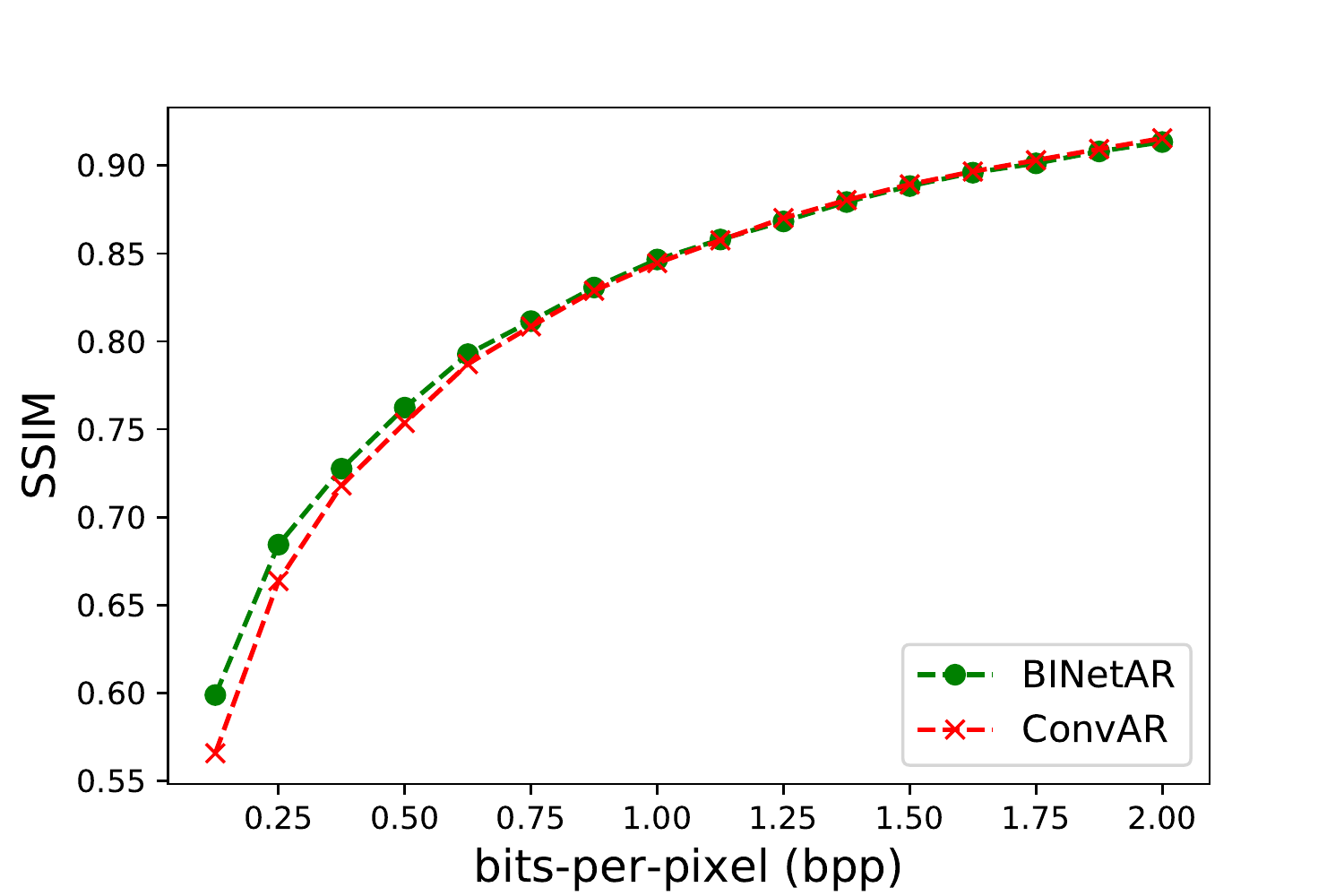}
		\label{fig:32_ssim_curve_ar}
	}
	\hfill
	\subfigure[PSNR]{
		\includegraphics[width=.45\textwidth]{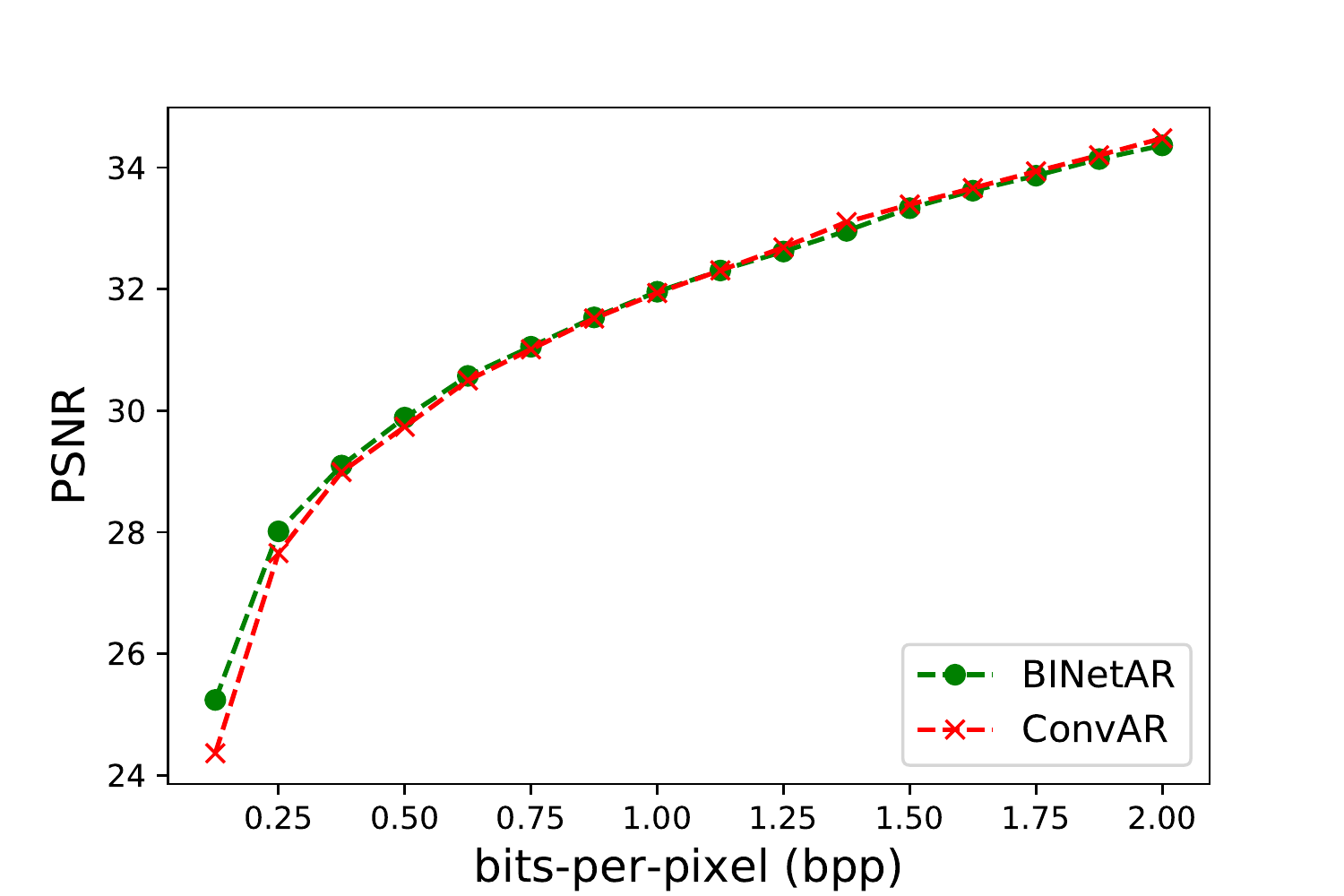}
		\label{fig:32_psnr_curve_ar}
	}
	
	\caption{
		BINetAR vs.\ ConvAR: rate-distortion curves for $32\times32$ image patches.
	}
\end{figure}

We first assess the abilities of BINetAR and ConvAR 
to reconstruct single $32\times32$
image patches centre-cropped from the test set.
Each model is trained to compress $32\times32$ patches,
for an intrinsic evaluation of model performance.
The resulting SSIM and PSNR rate-distortion curves are shown in
Figures~\ref{fig:32_ssim_curve_ar} and~\ref{fig:32_psnr_curve_ar}.
The variable bit rate is achieved by varying the number of encoding iterations from $I=1$ to $I=16$.
The figures indicate that at low bitrates close to the inpainting layer 
BINetAR gives a small but consistent improvement over ConvAR.

Dynamic bit assignment entails encoding different patch regions with varying bit allocations 
governed by a predetermined quality threshold such as PSNR.
This aids compression as image regions are not necessarily equally complex.
At low bitrates close to the inpainting layer 
BINetAR consistently produces patches of a higher quality than ConvAR.
This means that if dynamic bit assignment were implemented,
BINetAR would reach target quality thresholds after fewer encoding 
iterations (resulting in fewer bits) compared to ConvAR.

% Figure: 32x32 SSIM and PSNR curves

\subsubsection{Full Image Reconstruction}
The average PSNR and SSIM scores achieved by BINetAR and ConvAR on $224\times320$ test images
are compared in Table~\ref{tab:ssim_psnr_table}, for various bit allocations.
Although improvements are small, BINetAR consistently outperforms ConvAR across all bit depths. 
If one compares the first iteration of the two models,
BINetAR outperforms ConvAR by 8\% in terms of SSIM and 
results in a 3\% relative improvement in PSNR.
This comparison between the first iteration of the models is important 
as binary inpainting is only incorporated at the first stage of the BINetAR model.

% Table : SSIM and PSNR BINetAR vs ConvAR
\begin{table}[!h]
\caption{
	BINetAR vs.\ ConvAR: SSIM and PSNR scores at various bit-per-pixel (bpp) allocations for $224\times320$ images.
}
% \vspace{-3mm}
	\begin{center}
	% \small
		\begin{tabularx}{.95\linewidth}{@{}lCCCCCCCC@{}}
			\toprule
			\multirow{2}{*}{Model} &
			\multicolumn{3}{c}{SSIM} & \multicolumn{3}{c}{PSNR} \\
			\cmidrule{2-7}
				& \small{0.125~bpp} & \small{0.25~bpp} & \small{0.5~bpp} 
				& \small{0.125~bpp} & \small{0.25~bpp} & \small{0.5~bpp}\\
			\midrule
				ConvAR & $0.591$ & $0.712$  & $0.805$ & $22.486$ & $25.288$ & $27.488$ \\
				BINetAR & $\mathbf{0.639}$ & $\mathbf{0.732}$ & $\mathbf{0.813}$ & $\mathbf{23.222}$ & $\mathbf{25.643}$ & $\mathbf{27.623}$ \\
			\bottomrule
		\end{tabularx}
	\end{center}
\vspace{-3mm}
\label{tab:ssim_psnr_table}
\end{table}

\subsection{Quantitative Analysis: BINetOSR vs.\ ConvGRU-OSR}
\label{sec:binet_osr_eval}

Sixteen-iteration implementations of BINetOSR and ConvGRU-OSR are trained to assess the 
effect of incorporating a single inpainting stage on the performance of an OSR model.
Models are again evaluated on the CLIC test set~\cite{Freeman2018}.

\subsubsection{Patch Reconstruction}
We first asses BINetOSR's and ConvGRU-OSR's intrinsic capacity to reconstruct $32\times32$ 
patches center-cropped from the test data.
The resulting areas under 
the PSNR and SSIM rate-distortion curves are 
given in Table~\ref{tab:binet_osr_vs_conv_gru_auc}. 
Note that a greater area is indicative of increased perceptual quality 
across all sixteen allocated bitrates.
Table~\ref{tab:binet_osr_vs_conv_gru_auc} shows that incorporating learned inpainting 
into just one iteration of the ConvGRU-OSR model effectively increases its area under 
the PSNR and SSIM rate-distortion curves.\footnote{\textcolor{AN}{In Table~\ref{tab:binet_osr_vs_conv_gru_auc}, the area under the SSIM rate-distortion curve, which consists of multiple SSIM scores, is not necessarily bound to the range of an individual SSIM score $[-1, 1]$.}}

% Table : AUC BINetOSR vs ConvGRU (OSR)
\begin{table}[!h]
\caption{
	\textcolor{AN}{BINetOSR vs.\ ConvGRU-OSR: area under the curve for SSIM and PSNR rate-distortion,
	%plots
	calculated on $32\times32$ image patches.}
}
	\begin{center}
	% \small
		\begin{tabularx}{.6\linewidth}{lCC}
			\toprule
			\multirow{2}{*}{Model} &
				\multicolumn{2}{c}{Area under the curve} \\
			\cmidrule{2-3}
				 & SSIM & PSNR \\
			\midrule
				ConvGRU-OSR &  $1.661$ &  $64.352$ \\
				BINetOSR & $\mathbf{1.668}$ & $\mathbf{65.10}$\\
			\bottomrule
		\end{tabularx}
	\end{center}
\vspace{-3mm}
\label{tab:binet_osr_vs_conv_gru_auc}
\end{table}

\subsubsection{Full Image Reconstruction}
\begin{figure}[!h]
	\centering
% 	\vspace{-5mm}
	\vspace{-3mm}
	\subfigure[SSIM]{
		\includegraphics[width=.47\textwidth]{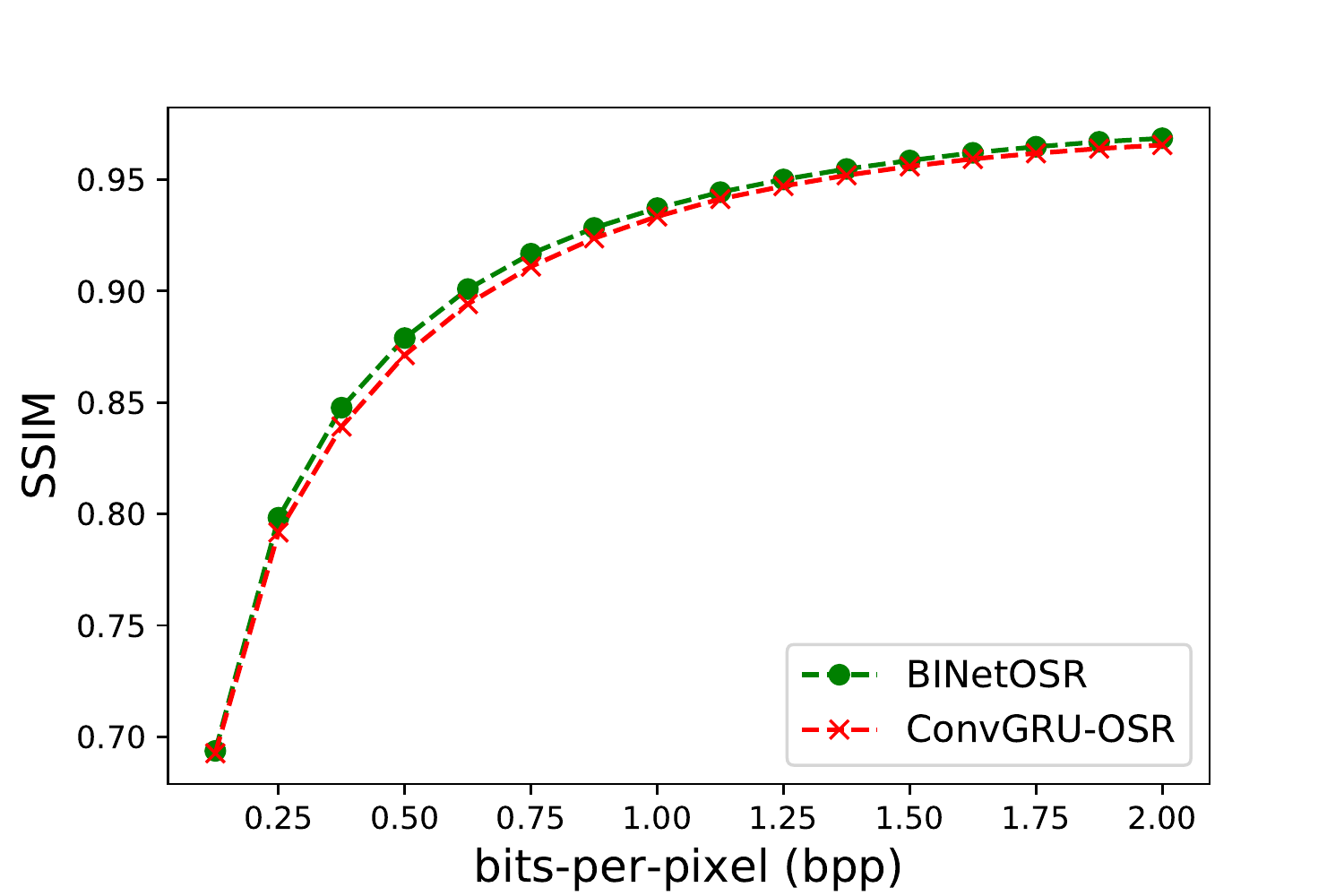}
		\label{fig:binet_osr_vs_conv_gru_ssim}
	}
	\hfill
	\subfigure[PSNR]{
		\includegraphics[width=.47\textwidth]{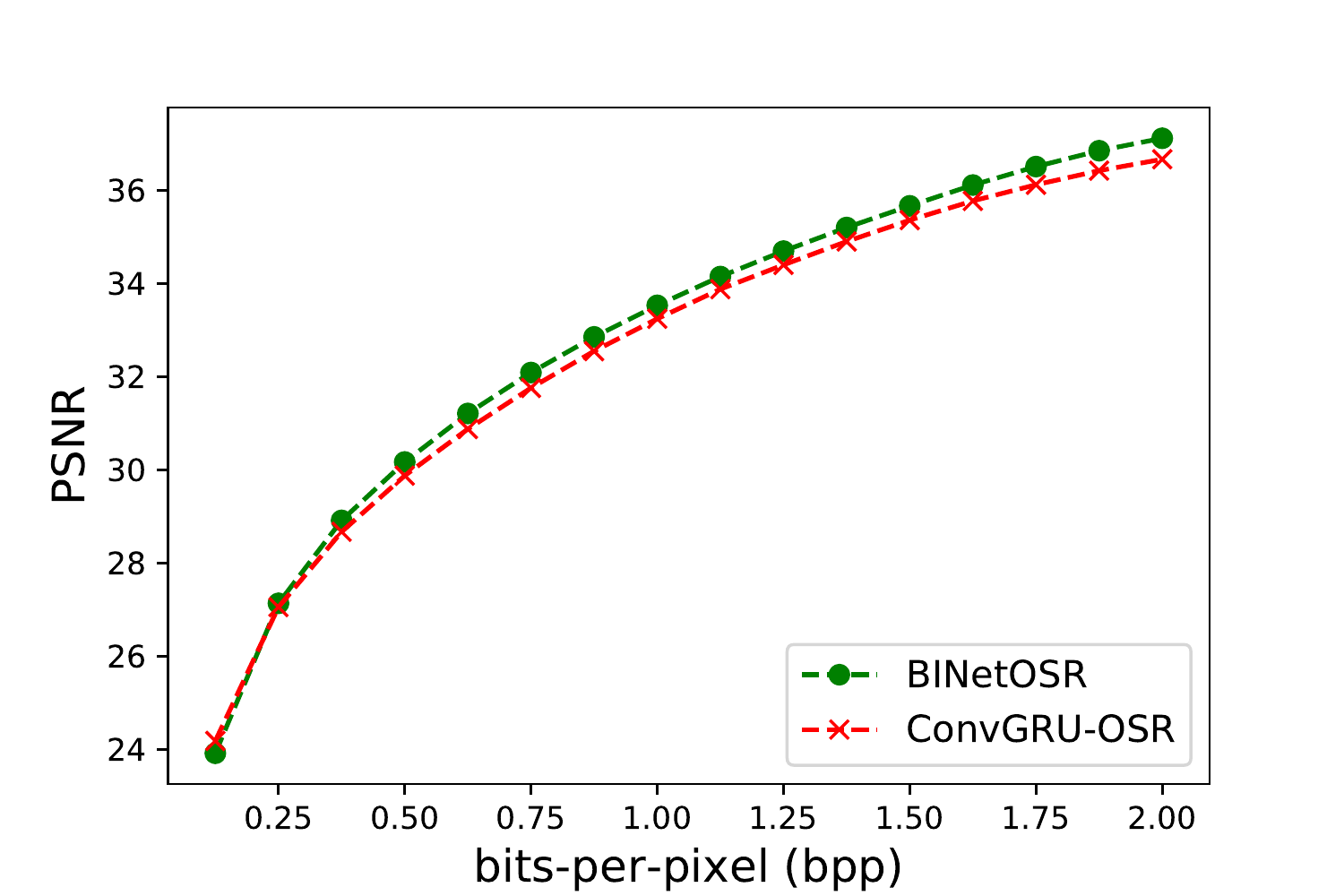}
		\label{fig:binet_osr_vs_conv_gru_psnr}
	}
	\caption{
		BINetOSR vs.\ ConvGRU-OSR: rate-distortion curves for complete $224\times320$ images.
	}
	\label{fig:rate_distorion_osr}
\end{figure}

The PSNR and SSIM curves achieved by BINetOSR and ConvGRU-OSR on $224\times320$ test images
are shown in Figures~\ref{fig:binet_osr_vs_conv_gru_psnr} and~\ref{fig:binet_osr_vs_conv_gru_ssim}.
Unlike the AR model, 
inpainting gains are more pronounced at stages further from the inpainting layer,
as recurrence allows BINetOSR to better propagate inpainting information 
to later decoding stages.
This forces the first stage to learn an inpainting strategy that is beneficial 
to the system as a whole as opposed to BINetAR where improvements 
are concentrated around the inpainting layer.
Again, while performance gains are small, they are consistent over the bit rates considered.

%
% Figure: BINetOSR vs ConvGRU (OSR) SSIM and PSNR Curves 224x320

\subsection{Quantitative Analysis: BINet vs.\ Standard Codecs}
\begin{figure}[!h]
	\centering
% 	\vspace{-5mm}
	\vspace{-3mm}
	\subfigure[CLIC]{
		\includegraphics[width=.475\textwidth]{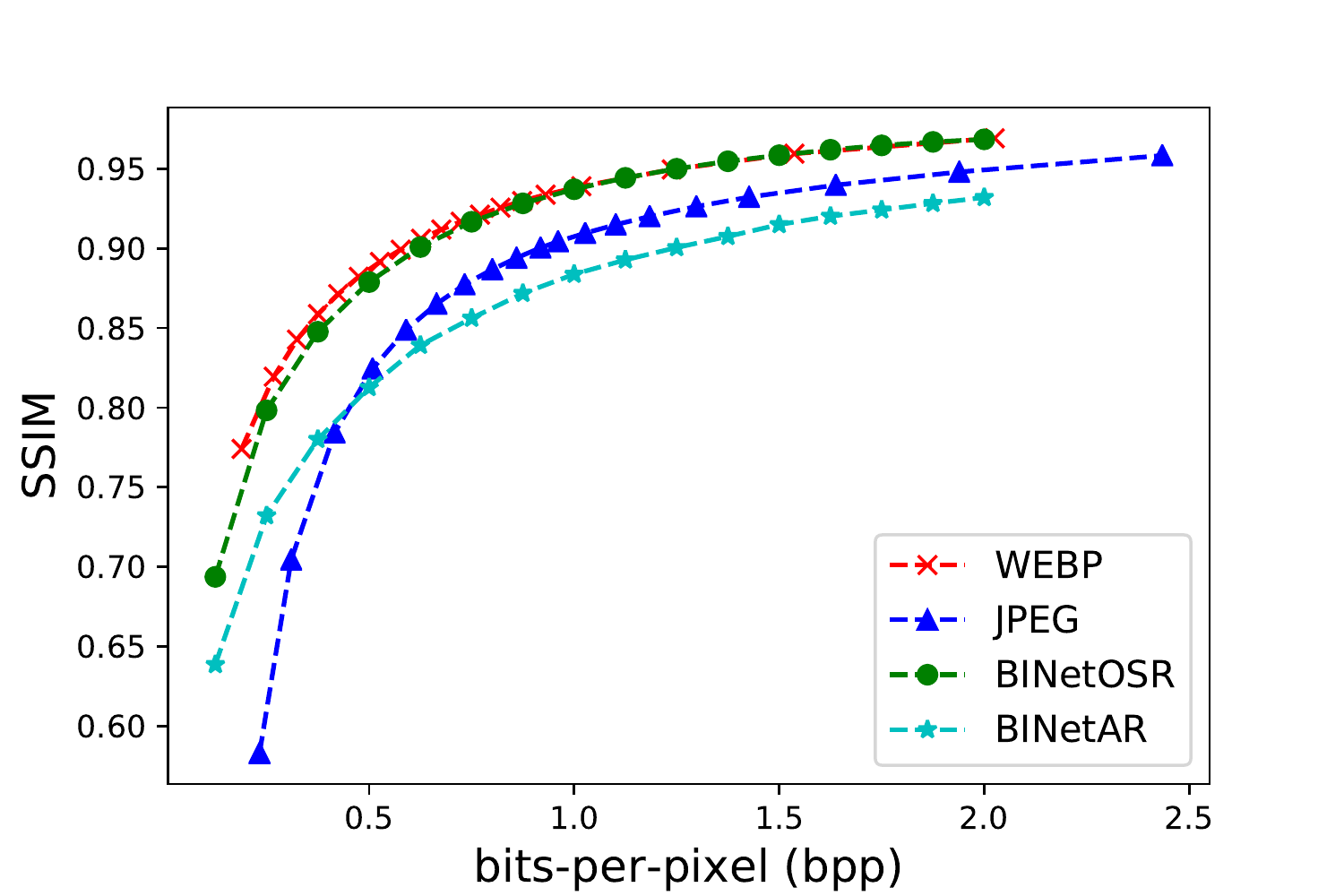}
		\label{fig:binet_vs_std_clic}
	}
	\hfill
	\subfigure[Kodak]{
		\includegraphics[width=.475\textwidth]{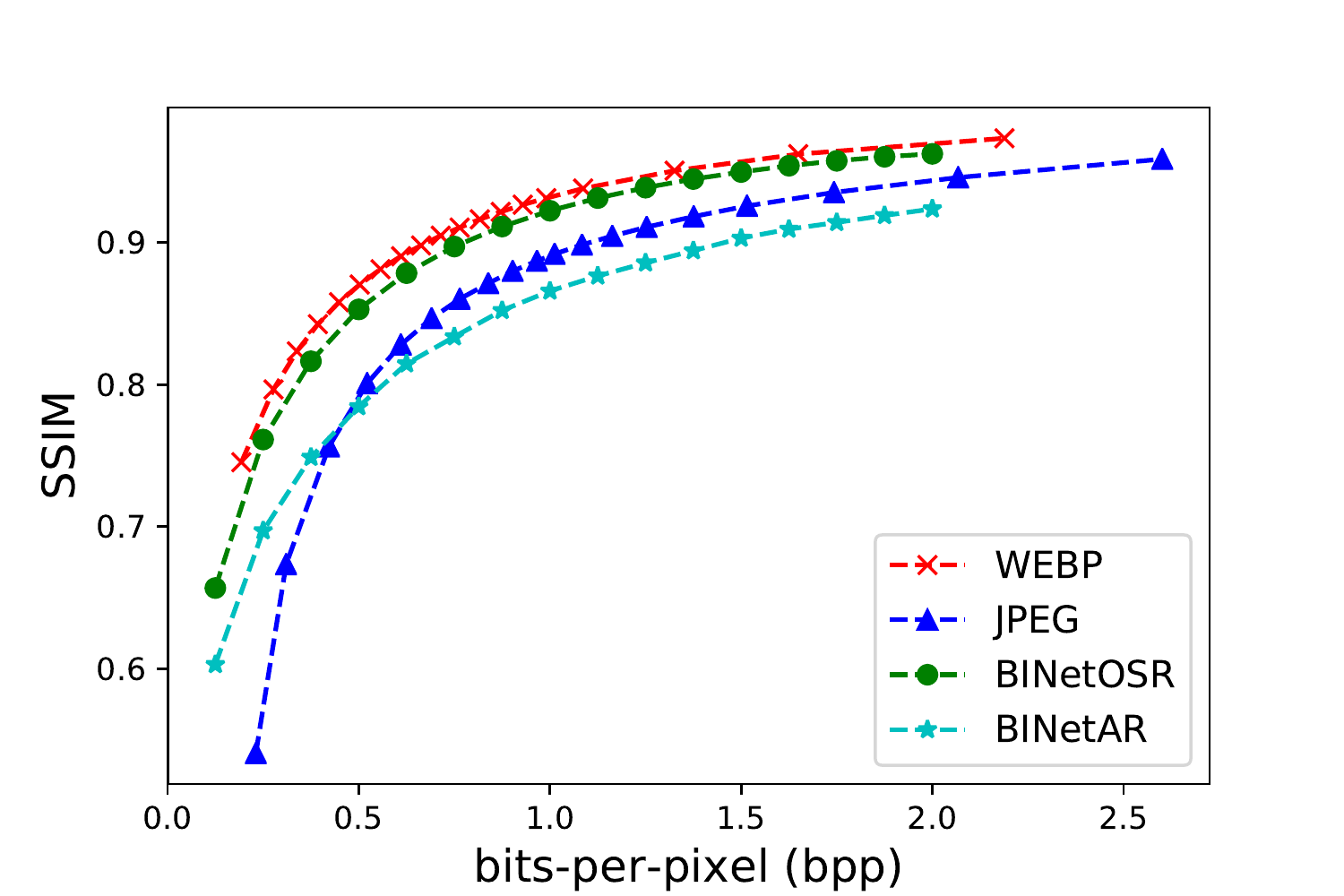}
		\label{fig:binet_vs_std_kodak}
	}
	\caption{
		\small{SSIM rate-distortion curves comparing BINet to standard image codecs on the CLIC Test~\cite{Freeman2018} and Kodak~\cite{Kodak1999} datasets.}
	}
	\label{fig:binet_vs_std_ssim}
	\vspace{-5mm}
\end{figure}

\begin{figure}[!h]
	\centering
	\subfigure[BINetAR]{
		\includegraphics[width=\textwidth]{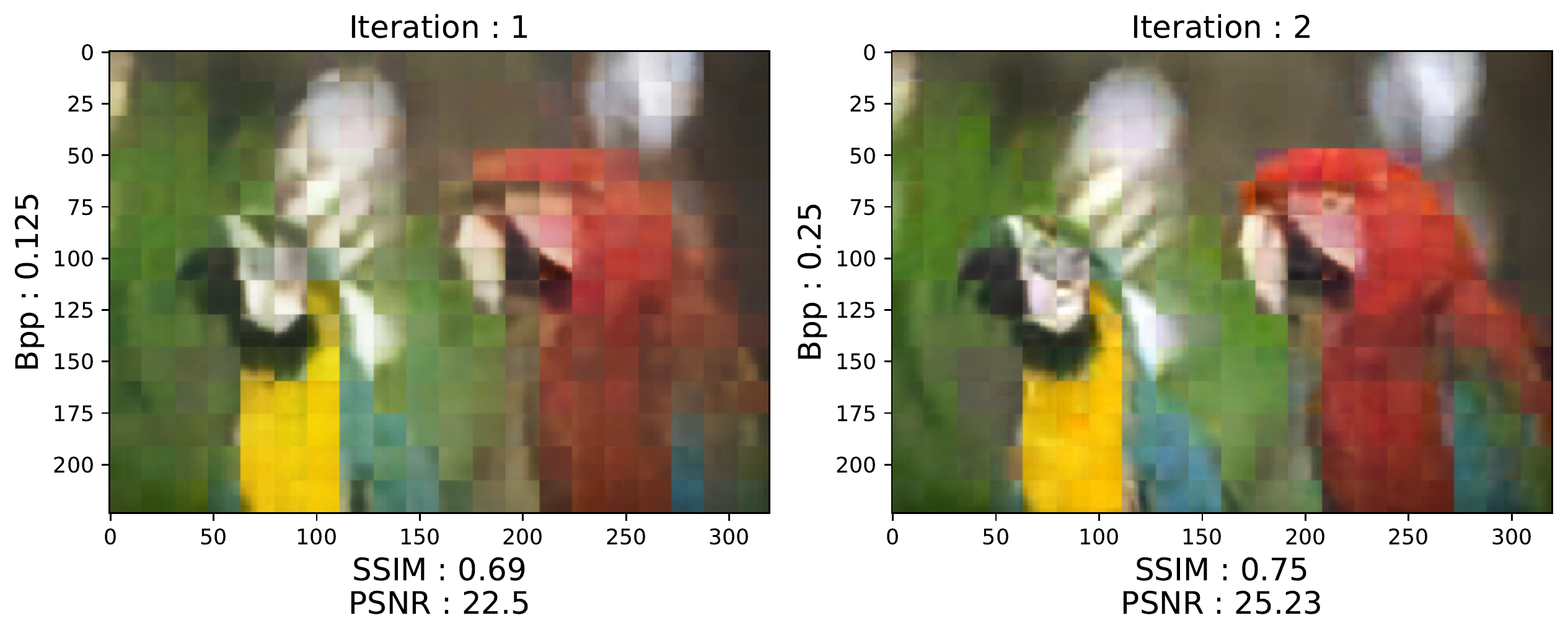}
	}
	\subfigure[ConvAR]{
		\includegraphics[width=\textwidth]{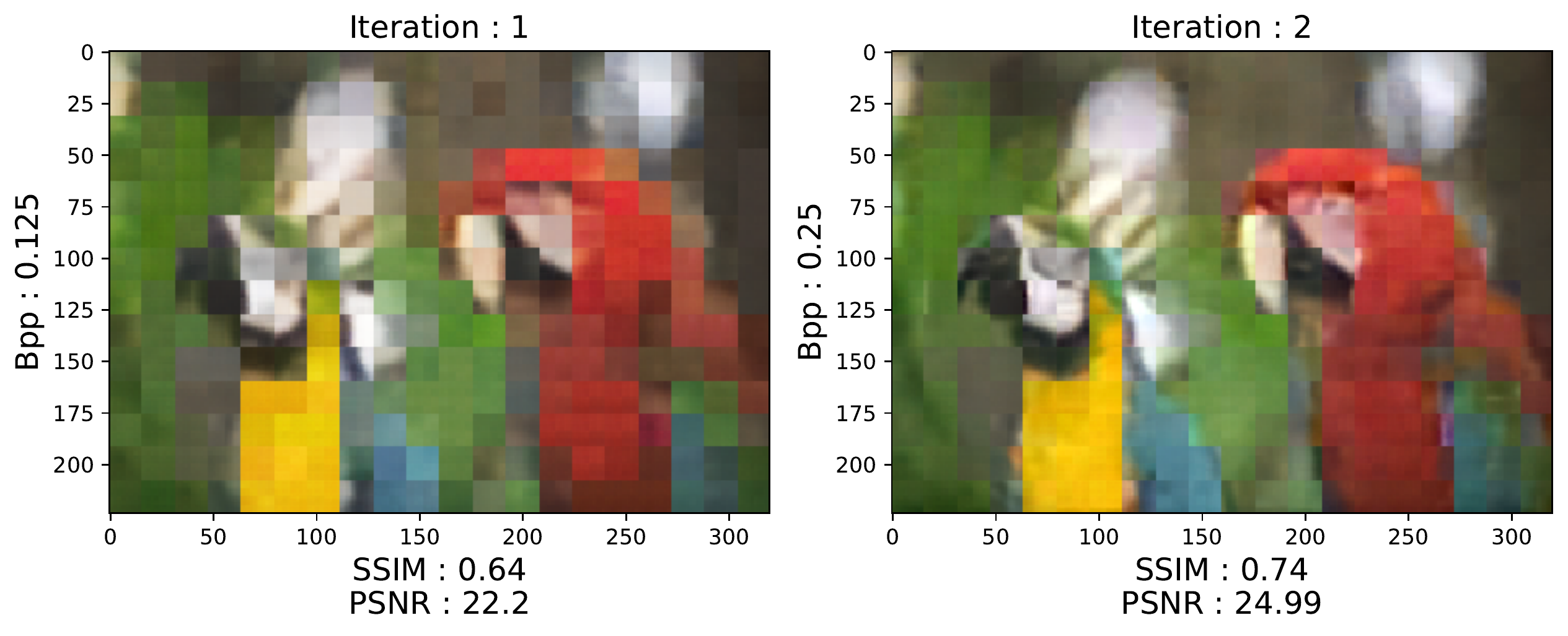}
	}
	\subfigure[JPEG]{
		\includegraphics[width=\textwidth]{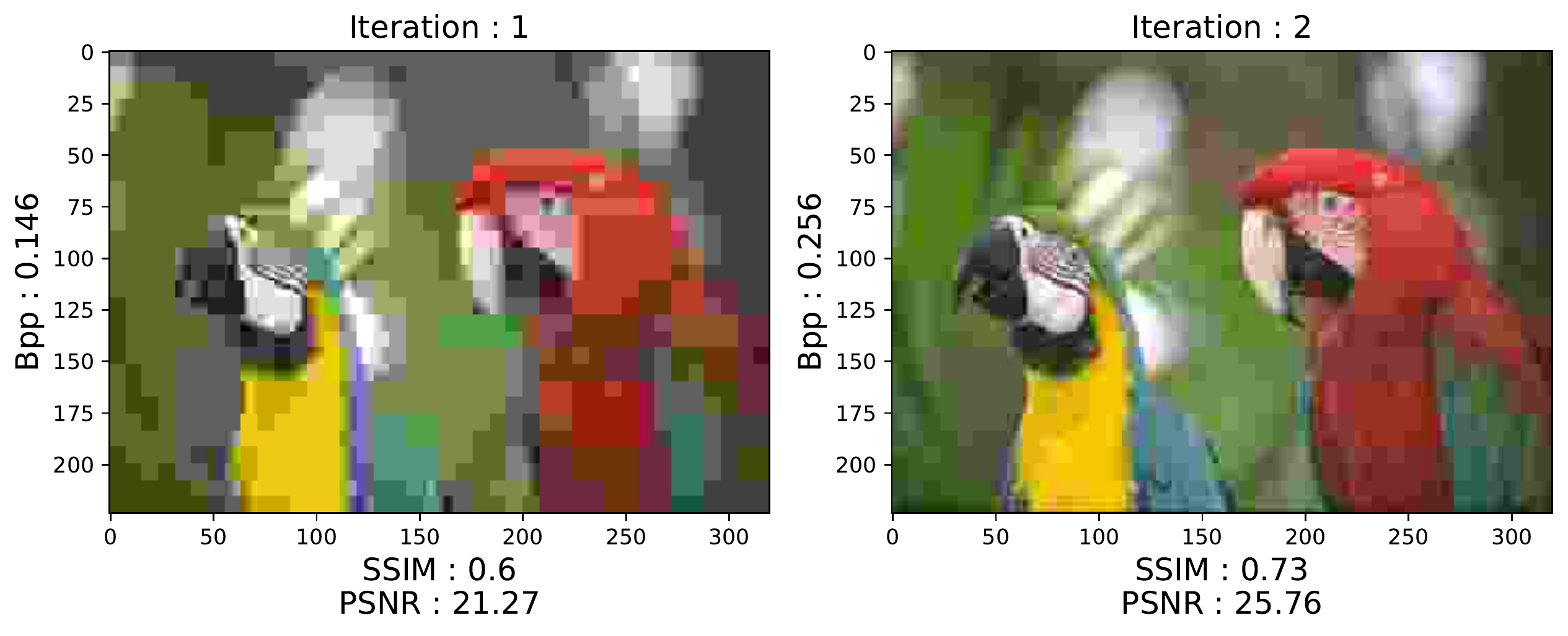}
	}
	\caption{BINetAR vs.\ ConvAR vs.\ JPEG: $224\times320$ image reconstructions.
	Image is from the Kodak dataset~\cite{Kodak1999}.}
	\label{fig:binet_ar_vs_jpeg_img}
\end{figure}

Up to now we focused on incorporating BINet 
into the ConvAR and ConvGRU-OSR approaches 
in order to investigate the effect of binary inpainting on 
patch-based compression in isolation.
Here we compare BINet to standard image codecs. 
Evaluation is carried out on full images from \textcolor{AN}{the CLIC Test~\cite{Freeman2018} and Kodak~\cite{Kodak1999} datasets} resized to $320\times224$ pixels.
Figure~\ref{fig:binet_vs_std_ssim} compares BINet's 
SSIM performance to that of WebP and JPEG. 
BINetAR outperforms JPEG at low bitrates, 
but gradually worsens at bitrates produced by stages further from the inpainting layer.
JPEG is patch-based, and Figure~\ref{fig:binet_ar_vs_jpeg_img} indicates 
that block artefacts arising through the use of an independent patch-based encoding scheme 
can be suppressed by BINetAR for enhanced quality at shallow bit allocations.

\begin{figure}[!h]
	\centering
	\subfigure[BINetOSR]{
		\includegraphics[width=\textwidth]{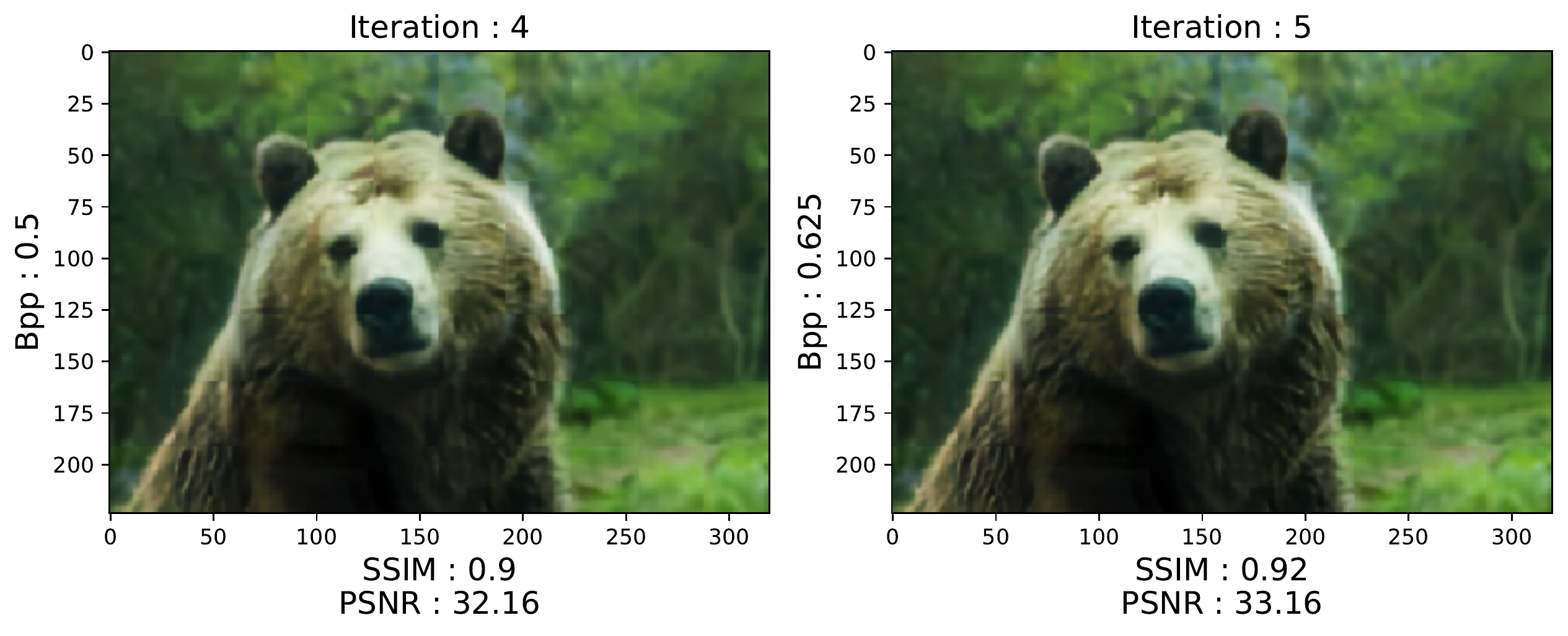}
	}
	\subfigure[ConvGRU-OSR]{
		\includegraphics[width=\textwidth]{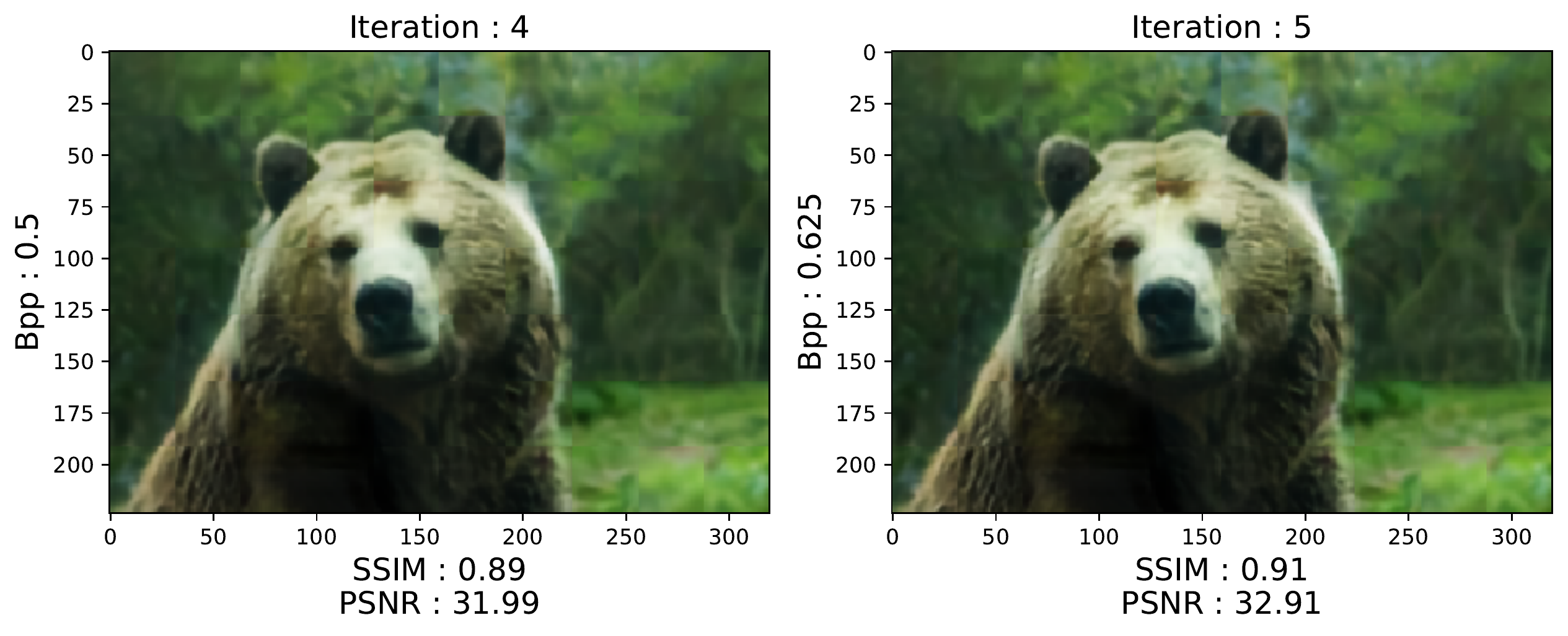}
	}
	\subfigure[WebP]{
		\includegraphics[width=\textwidth]{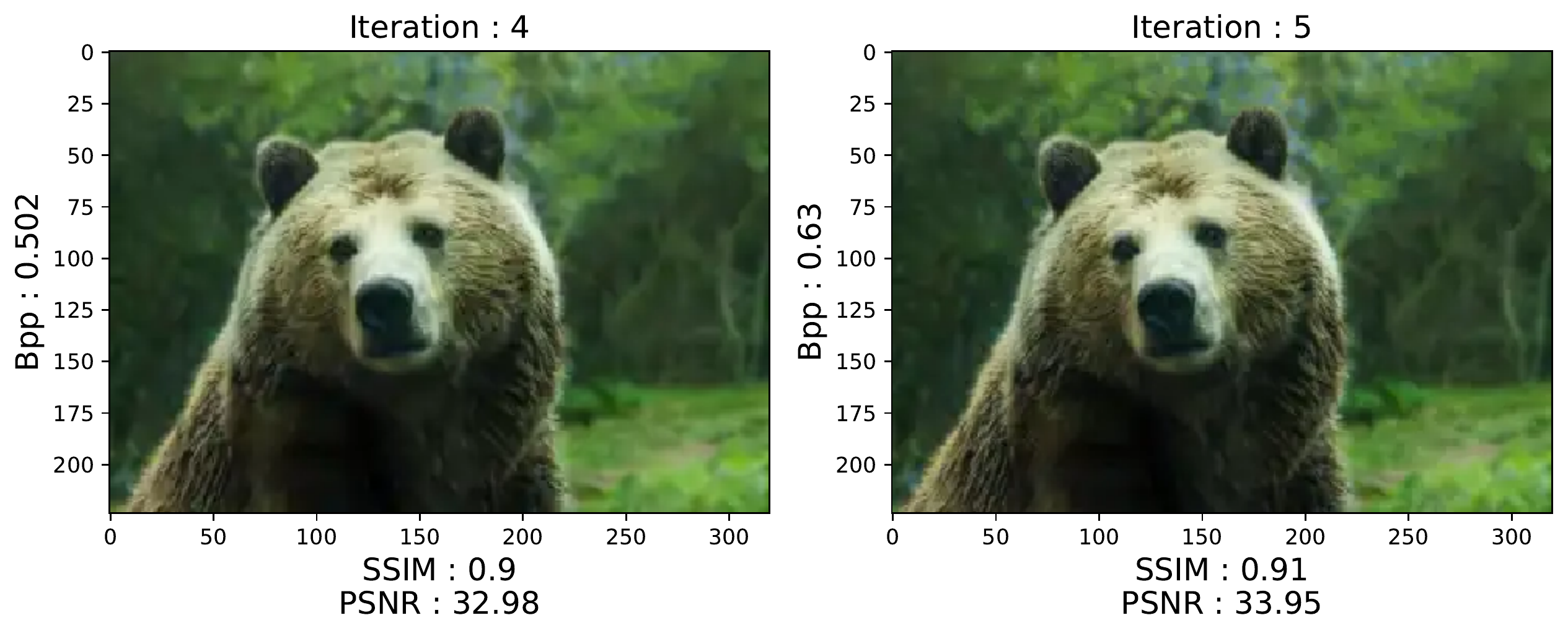}
	}
	\caption{\textcolor{AN}{BINetOSR vs.\ ConvGRU-OSR vs.\ WebP: $224\times320$ image reconstructions.
	Image is from the CLIC Test dataset~\cite{Freeman2018}.}}
	\label{fig:binet_osr_vs_webp_img}
\end{figure}

We showed in Section~\ref{sec:binet_experiments} that BINet is capable of learning 
more complex inpainting predictions than WebP.
Although the inclusion of a binary inpainting stage does consistently 
improve ConvGRU-OSR's performance (Figure~\ref{fig:rate_distorion_osr}), 
BINetOSR still falls short of outperforming WebP (\textcolor{AN}{Figures~\ref{fig:binet_vs_std_ssim} and~\ref{fig:binet_osr_vs_webp_img}}).
Using the same decoder module for inpainting and image patch reconstruction 
may result in a conflict between learning compression and the high quality inpainting 
shown in Section~\ref{sec:binet_experiments}.
One must also take into consideration that WebP and JPEG's codes 
are further compressed by additional compression techniques such as lossless entropy coding and dynamic bit assignment, whereas BINet's codes are not.

Our aim here was not to achieve state-of-the-art performance, 
but rather to investigate whether binary inpainting 
improves patch-based image compression in a deep neural network model, 
and this was shown in Sections~\ref{sec:binet_ar_eval} and~\ref{sec:binet_osr_eval}.
Any model can be used as the basis in the BINet framework, 
and future work will consider incorporating BINet into the 
more powerful recurrent models of~\cite{Johnston2018}.

%
% Figure: BINet vs WebP vs JPEG SSIM Curves Kodak and Clic Test Set 224x320

\begin{figure}[!h]
	\centering
	\vspace{-3mm}
	\subfigure[BINetAR]{
		\includegraphics[width=.475\textwidth]{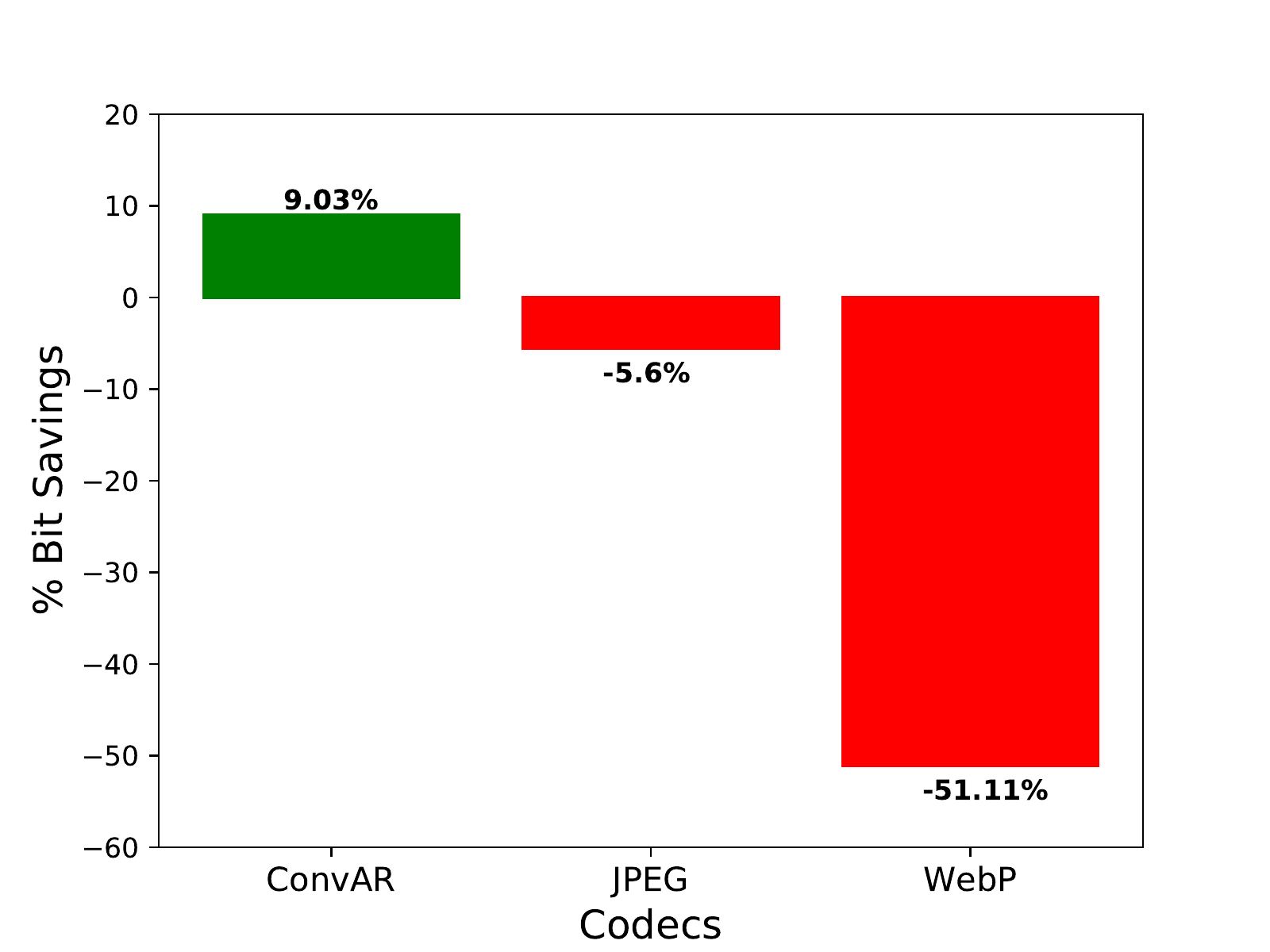}
		\label{fig:binet_ar_bd_rate_kodak}
	}
	\hfill
	\subfigure[BINetOSR]{
		\includegraphics[width=.475\textwidth]{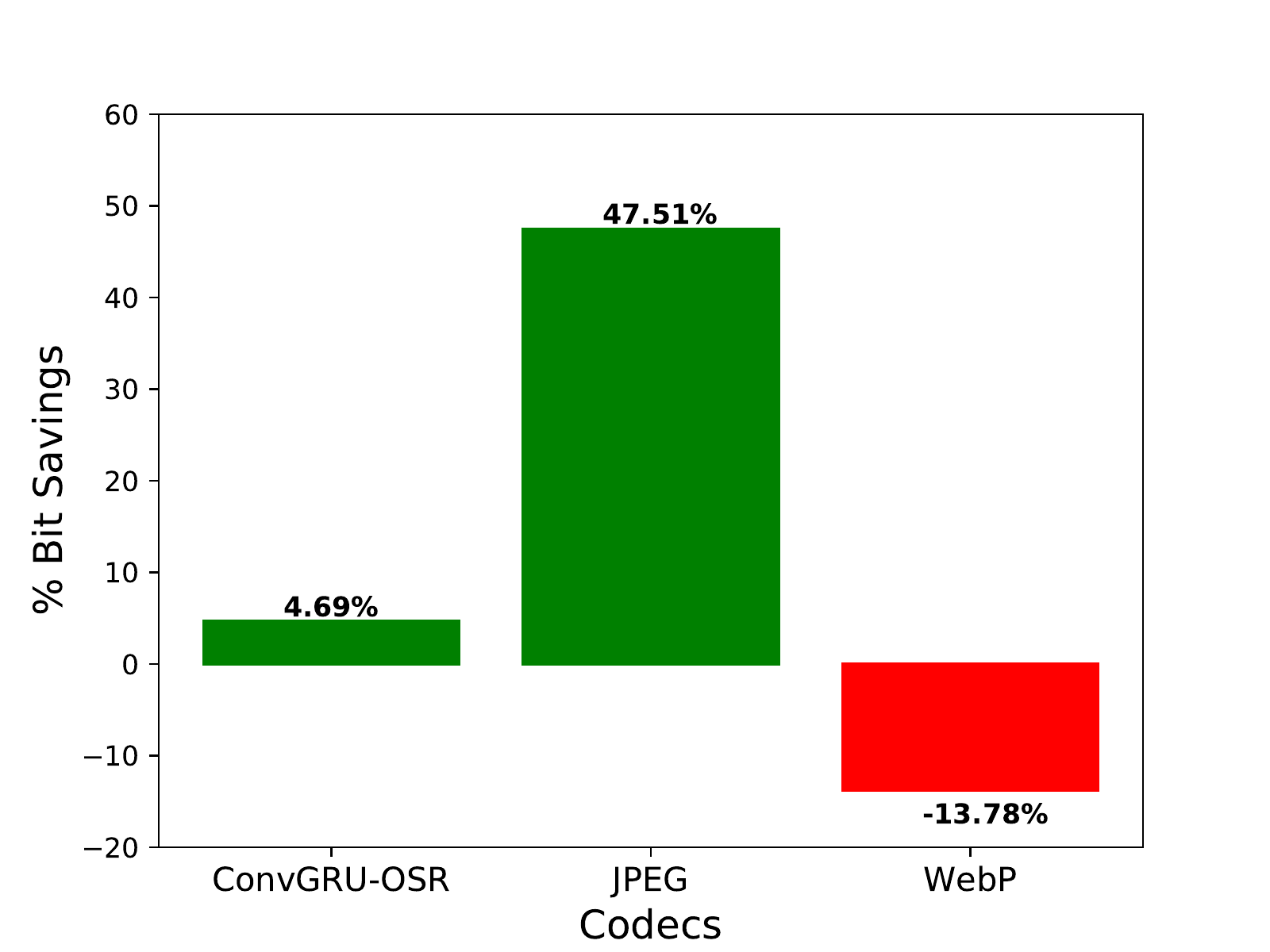}
		\label{fig:binet_osr_bd_rate_kodak}
	}
	\caption{
		\textcolor{AN}{SSIM BD-Rate plots showing the bit-savings afforded by BINetAR and BINetOSR on the Kodak~\cite{Kodak1999} datasets.}
	}
	\label{fig:bd_rate}
	\vspace{-3mm}
\end{figure}

\textcolor{AN}{\subsection{Summary of Results}}
% \textcolor{AN}{\section{Summary of Results}
\textcolor{AN}{
In summary,
%we plot SSIM BD-Rate curves~\cite{Bjontegaard1999}
\textcolor{AN}{we show SSIM BD-Rate plots~\cite{Bjontegaard1999}}
in Figure~\ref{fig:bd_rate} to illustrate the percentage bit-savings afforded by our binary inpainting codecs, BINetAR and BINetOSR, 
with respect to the baseline and standard image codecs considered. 
Each bar represents the average percentage bit-savings between BINet's SSIM rate-distortion curve and that of the codec named on the \textcolor{AN}{$x$}-axis.
Positive bit-savings are indicated by green bars, while red bars are used to signal instances where BINet falls short of outperforming a codec.
Most importantly, both plots in Figure~\ref{fig:bd_rate} indicate that BINet results in bit-savings compared to a baseline model without binary inpainting.}

%
% Figure: BD-Rate Plot w.r.t BINetOSR on Kodak and Clic Test Set 224x320
%

%
% Figure : 224 x 320 Image Reconstructions BINetAR vs. ConvAR vs. JPEG
%

%
% Figure : 224 x 320 Image Reconstructions BINetOSR vs. ConvGRU-OSR vs. WebP
%

% END EVALUATION

%% CONCLUSION
% START CONCLUSION

\section{Conclusion and Future Research}
% Conclusion
We introduced the Binary Inpainting Network (BINet), 
a novel framework that can be used to improve 
an existing system for patch-based image compression.
Building on ideas from image inpainting as well as deep image compression, 
BINet is novel in two particular ways.
Firstly, in contrast to work on inpainting,
BINet incorporates explicit binarisation in an encoder module, 
which allows it to be used for compression.
Secondly, in contrast to most deep compression models, 
BINet incorporates information from adjacent patches 
when decoding a particular patch.
The result is a patch-based compression method which
allows for parallelised inpainting from 
a full-context region without access to original image data.
In quantitative evaluations, we showed that BINet yields small 
but consistent improvements over baselines without inpainting. 
Qualitatively we showed that BINet results in fewer block artefacts at shallow bitrates 
compared to standard image codecs, 
resulting in smoother image reconstructions.

% Future Research
Apart from incorporating BINet into more advanced neural architectures\textcolor{AN}{, 
for example those containing deep entropy coding}, future work will also aim to explore alternative applications for binary inpainting such as
binary error correction and patch-based video-frame interpolation.
% END CONCLUSION

%%Harvard
\bibliographystyle{model1-num-names.bst}
\bibliography{my_content/my_refs}

\end{document}